\documentclass[
    10pt,
    onecolumn,
    secnumarabic,
    superscriptaddress,
    nobibnotes,
    nofootinbib,
    showkeys,
    aps,
    prd
]{revtex4-2}

\usepackage[T1]{fontenc}       % Better font encoding
\usepackage{amsmath, amssymb}  % Math symbols and environments
\usepackage{amsfonts}          % Extra math fonts
\usepackage{graphicx}          % Figures and graphics
\usepackage{bm}                % Bold math symbols
\usepackage{xcolor}            % Color support
\usepackage{comment}           % Commenting out blocks
\usepackage{hyperref}          % Hyperlinks
\usepackage{caption}
\usepackage{booktabs}
\usepackage{multirow}
\usepackage{graphicx}
\usepackage{subcaption}
\providecommand{\U}[1]{\protect\rule{.1in}{.1in}}

\usepackage{tocloft} % For customizing table of contents, figures, and tables

\providecommand{\U}[1]{\protect\rule{.1in}{.1in}}
\newcommand{\be}{\begin{equation}}
\newcommand{\ee}{\end{equation}}

\newcommand{\mincir}{\raise
-3.truept\hbox{\rlap{\hbox{$\sim$}}\raise4.truept\hbox{$<$}\ }}
\newcommand{\magcir}{\raise
-3.truept\hbox{\rlap{\hbox{$\sim$}}\raise4.truept\hbox{$>$}\ }}

\begin{document}
\title{Testing the Distance Duality Relation with Cosmological Observations at high Redshift using Artificial Neural Network}
\author{Yukang Xie}
\email{xieyk@gznu.edu.cn }
\affiliation{Guizhou Key Laboratory of Advanced Computing, Guizhou Normal University, Guiyang, 550025, China}

\author{Yang Liu}
\email{yangliu@pmo.ac.cn}
\affiliation{Purple Mountain Observatory, Chinese Academy of Sciences,No. 10 Yuanhua Road, Nanjing, 210023, China}

\author{Puxun Wu}
\thanks{Corresponding author}
\email{pxwu@hunnu.edu.cn}
\affiliation{Department of Physics and Synergistic Innovation Center for Quantum Effects and Applications, Hunan Normal University, Changsha, 410081, China}

\author{Xiangyun Fu}
\thanks{Corresponding author}
\email{xyfu@hnust.edu.cn}
\affiliation{Department of Physics, Key Laboratory of Intelligent Sensors and Advanced Sensor Materials, Hunan University of Science and Technology, Xiangtan, 411201, China}
\author{Nan Liang}
\thanks{Corresponding author}
\email{liangn@bnu.edu.cn}
\affiliation{Guizhou Key Laboratory of Advanced Computing, Guizhou Normal University, Guiyang, 550025, China}

\begin{abstract}
The cosmic Distance Duality Relation (DDR) is a fundamental prediction of metric gravity under photon number conservation. In this work, we perform a model-independent test of the DDR using Pantheon+ type Ia supernovae (SN Ia), \emph{Fermi} gamma-ray bursts (GRBs) with the FULL and GOLD samples, the Dark Energy Spectroscopic Instrument (DESI)  Data Release 2 (DR2) baryon acoustic oscillation (BAO) measurements, and the galaxy-scale strong gravitational lensing (SGL) system samples at high redshift $0.01 < z \lesssim 8$ using an  artificial neural network (ANN) approach.
%We adopt two commonly used parameterizations for possible DDR deviations: the $P_1$ model ($\eta(z) = 1 + \eta_0 z$) and the $P_2$ model ($\eta(z) = 1 + \eta_0 z/(1+z)$).
%Three fitting schemes are considered: (i) all three parameters ($r_{\rm d}$, $M_{\rm B}$, and $\eta_0$) are free; (ii) $r_{\rm d}$ is fixed at 147.78~Mpc, while $M_{\rm B}$ and $\eta_0$ are free; and (iii) both $r_{\rm d}$ and $M_{\rm B}$ are fixed, adopting external calibrations $M_{\rm B}^{\rm M24}$ or $M_{\rm B}^{\rm SH0ES}$, leaving only $\eta_0$ as a free parameter.
Our results show that %for both $P_1$ and $P_2$ models and under all fitting schemes,
the standard DDR %($\eta(z) = 1$)
is consistent with cosmological observations at high redshift within the $\sim 2 \sigma$ confidence level. %, with no significant
\end{abstract}
\maketitle

\section{Introduction}
The cosmic Distance Duality Relation (DDR), also known as Etherington's Reciprocity Theorem \citep{Etherington1933}, establishes a universal link between the luminosity distance (LD, $D_L$) and the angular diameter distance (ADD, $D_A$), which can be expressed mathematically as
\begin{equation}
	\eta(z) \equiv \frac{D_{\rm L}(z)}{(1+z)^2 D_{\rm A}(z)} = 1
	\label{eq:cddr}
\end{equation}
This relation holds strictly in any cosmological model based on metric theories of gravity and Lorentzian (pseudo-Riemannian) spacetime geometry \citep{Ellis2007}.
The validity of the DDR relies on three fundamental physical assumptions:
(i) gravity is described by a metric theory (such as general relativity);
(ii) photon number is conserved during propagation (i.e., the Universe is completely transparent, with no photon decay, creation, or absorption); and
(iii) photons travel along unique null geodesics.
The DDR forms a cornerstone of modern observational cosmology, providing a model-independent consistency test linking distance scales inferred from different observational methods \citep{Lima2003}.
The tests of the DDR have become an important probe for exploring new physics. Any statistically significant deviation %from $\eta(z) = 1$
may reveal unmodeled systematic errors or point to phenomena beyond the standard cosmological framework.
%\textbf{For instance, in certain modified gravity theories, photons may not propagate strictly along null geodesics; in scenarios with non-minimal photon-scalar field coupling, or through mechanisms such as photon-axion oscillations or
%intergalactic dust opacity, the DDR could be violated \citep{Avgoustidis2010,Li2011,Cardone2012}.}
For instance, in certain modified gravity theories, photons may not propagate strictly along null geodesics, leading to deviations from the standard distance relations. Moreover, processes that violate photon number conservation; such as non-minimal photon scalar field couplings, photon axion oscillations, or absorption and scattering by intergalactic dust can systematically affect the observed luminosity distance. In these scenarios, %the fundamental assumptions underlying the DDR are no longer satisfied, and
the DDR  could be violated \citep{Avgoustidis2010, Li2011, Cardone2012}.

DDR can be tested with observational data, e.g., the LD from type Ia supernovae (SNe Ia) and the ADD from galaxy clusters combined Sunyaev-Zel'dovich (SZ) and X-ray cluster observations. %surface brightness measurements under the assumption of hydrostatic equilibrium \citep{DeFilippis2005,Bonamente_2006}. %baryon acoustic oscillations (BAO), and strong gravitational lensing (SGL).
Due to the differences in redshift coverage and sampling density among various probes, directly comparing $D_{\rm L}(z)$ with SNe Ia and $D_{\rm A}(z)$ with the galaxy clusters often faces the challenge of redshift mismatch.
Earlier studies addressed this issue by selecting SNe Ia whose redshifts are closest to the galaxy clusters within a small interval  or
binning SNe Ia data within such a small interval \citep{Holanda2010,Li2011,Meng2012}.

In order to overcome redshift mismatch biases and preserve the integrity of observational samples,
\citet{Liang2013} introduced a new consistent method which derives LDs of a certain SN Ia point at the same redshift of the corresponding galaxy cluster, %by interpolating from the nearby SNe Ia,
effectively relating luminosity distances to the ADDs without cosmological assumptions.
Recently, %nonparametric techniques have been used to test DDR, e.g.,
Gaussian process (GP) \citep{Seikel2012} %in which the reconstruct only depends on the chosen kernel function and the data observed
have been tested the possible violation of DDR  with the available local data %including SNe Ia, galaxy clusters, the cosmic chronometer (CC) Hubble parameter, and BAO
in many  works \citep{Lin2018,Zheng2020,Mukherjee2021,Bora2021,Kumar2022,Wang_2024}. %Kumar D., Rana A., Jain D., Mahajan S.,
%Non-parametric reconstruction techniques, such as Gaussian Processes (GPs) and
%However, the results of GP become unreliable with large uncertainties at high redshift where data are sparse \citep{Lin2018}.
%Moreover, the choice of kernel function influences the reconstruction results, and selecting an appropriate kernel often relies on subjective experience \citep{Bernardo2021,Zhang2023,johnson2025}.
More recently, researchers have explored their applications in astronomy to tackle complex data challenges with the development of deep learning techniques.
\citet{Wang2020} used an artificial neural networks to reconstruct functions (ReFANN) from observational data.
Several studies have applied ReFANN to test the DDR \citep{Liu2021,Xu2022,Yang2025a,Yang2025b}.

%For the observational data to test DDR,
%, including Type Ia supernovae (SNe Ia), baryon acoustic oscillations (BAO), strong gravitational lensing time delays, and combined Sunyaev-Zel'dovich (SZ) and X-ray cluster observations.% to place increasingly stringent constraints on $\eta(z)$.
%Achieving high-precision DDR tests with the observational data requires accurate and mutually independent measurements of $D_{\rm L}(z)$ and $D_{\rm A}(z)$ over a broad redshift range.
%At low to intermediate redshifts $z \lesssim 2$, SNe Ia remain the most reliable probes of luminosity distance.
%The Pantheon+ sample \citep{Scolnic_2022}, containing 1701 spectroscopically confirmed SNe Ia within $0.01 < z < 2.3$, represents the most precise set of apparent magnitude measurements currently available.
%Achieving %high-precision
DDR tests with the observational data requires accurate and mutually independent measurements of $D_{\rm L}(z)$ and $D_{\rm A}(z)$ over a broad redshift range. The galaxy clusters sample can give the angular diameter distances only at $z<0.9$ \citep{DeFilippis2005,Bonamente_2006}.
Recently, many works used ADDs obtained through other measurements to test
the DDR: e.g.,
%Galaxy clusters-combining SZ effect and X-ray surface brightness measurements under the assumption of hydrostatic equilibrium \citep{Bonamente_2006,Wan2021,Avila_2025};
the gas mass fraction measurements in galaxy clusters \citep{Holanda2012,Wang2013,Holanda2022};
baryon acoustic oscillations (BAO) measurements of the angular scale of the acoustic peaks from the latest large-scale structure surveys \citep{Cardone2012,Wu2015,Ma2018,Lin2018,Xu2022,Zheng_2025},  %such as SDSS and DESI \citep{DESI2024a,DESI2024b,DESI2025a,DESI2025b}.
the ADD ratio from strong gravitational lensing (SGL) systems  \citep{Liao2016,Liao2019,Zhou2021,Lima2021,Holanda2022,Tang2023},
the time-delay distance $D_{\Delta t}$ derived from  SGL \citep{Wong2020,Shajib2023,Birrer2024,Gahlaut2025,Tang2025},
as well as  the compact radio quasars \citep{Zheng2020,Liu2021,Yang2025b}. %which is directly related to $D_{\rm A}$  %; and (iv) H II galaxies—using the empirical relation between $H\beta$ luminosity and the ionized gas velocity dispersion as a distance indicator \citep{Chavez2014,GonzalezMoran2021,Oh2022,Zheng_2025,Wei2025}.
%Current precision has reached the few-percent or even sub-percent level \citep{Alam2021_eBOSS,Holanda2020,Collett2019,Liu2021}.
For the observational data of LDs,
SNe Ia remain the most reliable probes  at $z \lesssim 2$ \cite{Scolnic2018,Scolnic2022}.
Other astrophysical sources of LDs for testing
the DDR have been proposed, e.g., the observational Hubble data (OHD) \cite{Avgoustidis2010,Chen2016},
the gravitational wave (GW) \cite{Fu2019,Liao2019,Lin2021,Huang2025a}, % “standard sirens”,
quasars \citep{Zheng2020,Avila_2025},
and H II galaxies \cite{Liu2021,Zheng_2025}. %, which using the empirical relation between $H\beta$ luminosity and the ionized gas velocity dispersion as a distance indicator \citep{Wei2025};

Moreover, gamma-ray bursts (GRBs) can serve as valuable complementary probes due to the extremely high intrinsic luminosities %at higher redshifts ($z \gtrsim 1.4,\ \text{extending up to}\ z \sim 9$),
%Although GRBs exhibit large intrinsic scatter,
%with the empirical multi-variable relations of GRBs, %-such as the Amati relation (between the isotropic-equivalent energy $E_{iso}$ and the spectral peak energy $E_{peak}$)-
which allow their standardization as potential standard candles at high-redshift extending up to $ z \sim 9$ \citep{Amati2002,Ghirlanda2004,Yonetoku2004,LZ2005,LZ2006,Schaefer2007}.
%In order to avoid the circularity problem, %\cite{Liang2008} proposed
Model independent methods to calibrate GRBs  %at low-redshift
from  SNe Ia \citep{Liang2008}, OHD %derived from the cosmic chronometers (CC) method
\citep{Amati2019}, quasars \citep{Dai2021}, and %the angular diameter distances of
galaxy clusters \citep{GD2022}, as well as the simultaneous fitting method \citep{Amati_2008} have been proposed to avoid the circularity problem; % and used a B\'{e}zier parametric curve to calibrate the GRB relation.
therefore GRBs can be proposed for cosmological use %without assuming a prior cosmological model
%\citep{Liang2010,Liang2011,WL2010,Wang2011,Gao2012,Demianski_2021,Khadka_2021,Luongo2021,Luongo2025a,Luongo2025b,Liu2022,Liang2022,LZL2023,Favale2024,Dainotti2024,Cao2024,Cao2025,Bargiacchi2025,Alfano2025,Luongo2025,Colgain2026}.
\cite{Liang2010,Liang2011,WL2010,Wang2011,Gao2012,Demianski2017,Khadka_2021,Luongo2021,Luongo2025a,Luongo2025b,Liu2022,Liang2022,LZL2023,Favale2024,Dainotti2024,Cao2024,Cao2025,Bargiacchi2025,Alfano2025,Luongo2025,Colgain2025}.
%\cite{Liang2008} proposed a model-independent calibration method by interpolating GRB data from low-redshift SNe Ia observations; therefore GRBs can be used to constrain cosmological parameters without assuming a prior cosmological model \citep{Liang2010,Liang2011,Wang2011,Amati_2008,Amati2019,Demianski_2021,Khadka_2021,Liu2022}.

Recently, GRB data  %can be used to
combined with SNe Ia
have been used to test DDR \citep{Fu2017} and probe cosmic opacity \citep{Holanda2014,Hu2017,Holanda2018}  at high redshift.
\citet{Alfano2026} tested the DDR by combining %low- and intermediate/high-z data sets, such as
OHD, galaxy clusters, the second data release (DR2) from the DESI Collaboration, Pantheon, and GRBs with the $E_{\rm iso}-E_{\rm p}$ or $L_{\rm 0}-E_{\rm p}-T$ relations.

More recently,
%\cite{Yang_2025} explored the impact of calibration priors (e.g., SNe absolute magnitude $M_{\rm B}$), %or QSO linear size $l$), showing that careful treatment is necessary to avoid apparent DDR violation.
\citet{Teixeira2025}  constrain five phenomenological parameterisations of DDR violation using BAOs from the Dark Energy Spectroscopic Instrument (DESI) survey %calibrated with the sound horizon derived from Planck Cosmic Microwave Background data
and the SN Ia catalogue. %calibrated with the SN absolute magnitude from SH0ES.
Li et al.\cite{Li2025} conducted a comprehensive test of the DDR by combining BAOs from the SDSS and DESI  with SN data. %from Pantheon+ and DESY5.
%\cite{Avila_2025} tested the validity of the DDR by combining data from transverse BAOs, galaxy clusters, Pantheon+ sample and from quasar catalogs.
Lopez-Hernandez et al.\cite{Lopez-Hernandez2025} performed a consistency check of DESI DR2 BAO constraints by reconstructing %the same quantities
from DES SNe in bins with the same effective redshift to confirm systematics in either DESI BAO or DES SNe.
%\cite{Poulin2024,Shah2024a,Shah2024b,Kanodia2026}
%Kanodia et al. \cite{Kanodia2026} tested the validity of the DDR by  combining the Megamaser Cosmology Project with the Pantheon + sample  at very low redshifts ($z<0.04$),  and  DESI DR2 in combination with SNIa data at high-redshift, highlighting the critical role of the early-($r_d$) - late ($M_B$) calibration in testing the DDR using these two probes.}
Wang et al. \cite{Wang2025} tested the DDR through a novel, model-independent method inspired by the two-point diagnostic approach with DES-SN 5YR and Pantheon sample reconstructed using the ANN technique. This methodology effectively eliminates all nuisance parameters, including the sound horizon scale $r_{\rm d}$ from BAO and the absolute magnitude $M_B$ from SN Ia.
Keil et al.\cite{Keil2025} combined the latest observational data (such as DR1, Pantheon+, SH0ES Cepheid calibrations, and DES-SN5YR) to perform a rigorous test of the DDR by a parametrised approach and also use model-independent Generic Algorithms (GA), which are a machine learning method where functions evolve loosely based on biological evolution.
Dinda et al.\cite{Dinda2025} performed a calibration-independent and model-agnostic consistency test between DESI DR2 BAO and multiple SNIa datasets.
Kanodia et al.\cite{Kanodia2026} tested the validity of the DDR by  combining the Megamaser Cosmology Project with the Pantheon + sample  at very low redshifts ($z<0.04$),  and  DESI DR2 in combination with SNIa data at high-redshift, highlighting the critical role of the early ($r_{\rm d}$) - late ($M_{\rm B}$) calibration in testing the DDR using these two probes.

In our recent work,
%In this paper, we independently reconstruct the luminosity distances of the Pantheon+ SNe Ia sample and the high-redshift GRB sample using an artificial neural network (ANN) approach, and combine them with the angular diameter distance measurements from DESI DR2 BAO data to perform a systematic test of the distance duality relation (DDR).
%We adopt two absolute magnitude priors:
%$M_{\rm B}^{\rm M24} = -19.353^{+0.073}_{-0.078}~\mathrm{mag}$ (from \cite{Mukherjee_2024}, based on CC and Pantheon+), and
%$M_{\rm B}^{\rm SH0ES} =  -19.253 \pm 0.027~\mathrm{mag}$ (from \cite{Riess2022}, based on Pantheon+).
%We consider two different parameterizations of $\eta(z)$, namely, i) P1:$\ \eta_{\rm th}(z;\eta_0) = 1 + \eta_0 z$ \citep{Bassett2004}, and ii)P2: $\ \eta_{\rm th}(z;\eta_0) = 1 + \eta_0 \frac{z}{1+z}$ \citep{Holanda2010}. In addition, as a comparison, we also perform a non-parametric test of the DDR.
% --- %
%In our previous work,
\citet{WangLiang2024}  presented a sample of long GRBs  from 15 years of the Fermi-GBM catalogue with identified redshift, in which the GOLD sample contains 123 long GRBs at $z\le5.6$ and the FULL sample contains 151 long GRBs with redshifts at $z\le8.2$.
\citet{Zhu2025} constrained the phenomenological interacting dark energy model with \emph{Fermi} GRBs and DESI DR2.
\citet{Huang2025b} employed an ANN framework to
% reconstructe
reconstruct %in a model-independent way
by considering the physical correlations in the data with the covariance matrix and KL (Kullback-Leibler) divergence into the loss function and calibrate the Amati relation. %  by selecting the optimal ANN model.
\citet{Luo2025} proposed a new method, termed Neural Kernel Gaussian Process Regression (NKGPR), to reconstruct the SNe Ia dataset to test the DDR.

In this work, we test the DDR using DESI DR2, SGL data together with GRB data and the Pantheon+ SN~Ia sample using ANN approach.
%This combination enables both parametric and non-parametric determinations of $\eta(z)$ based on reconstructed luminosity and angular-diameter distances.
The structure of the paper is as follows. Section~\ref{sec:data} presents the data, %and methodology,
including GRB samples, DESI DR2 BAO data, SGL systems, and SN~Ia datasets. %the ANN reconstruction, and the adopted DDR parameterizations.
Section~\ref{sec:recon} describes reconstructions from SN~Ia and GRBs. Section~\ref{sec:test_ddr} reports the DDR tests and the corresponding results. In Section~\ref{sec:conclu}, we present
the  discussions and conclusions.

\section{Data
}\label{sec:data}
In this study, we use the latest BAO data from DESI DR2 \citep{DESI2025a,DESI2025b}, SGL data \citep{Chen2019,Amante2020}, the \emph{Fermi} GRB sample \cite{WangLiang2024}, as well as the Pantheon+ SNe Ia sample \citep{Scolnic2022}, and then show our methodology adopted for DDR validation.

\subsection{GRBs Samples}
In this work, we employ the updated sample of GRBs from the \emph{Fermi} satellite \citep{WangLiang2024}, in which the GOLD sample contains 123 long GRBs at $z\le5.6$ and the FULL sample contains 151 long GRBs with redshifts at $z\le8.2$. The \emph{Fermi}-GBM catalogue at low-redshift can be calibrated from the latest OHD  with the cosmic chronometers method by using a Gaussian Process to obtained GRBs at high-redshift $z\ge1.4$.
%Unlike type Ia supernovae,
Therefore, GRBs can effectively avoid the systematic uncertainties associated with SNe calibration of the absolute magnitude $M_{\rm B}$, which directly provide the distance modulus $\mu$ without relying on the value of $M_{\rm B}$.
The number of \emph{Fermi} GRBs at $z\ge1.4$: $N_{\mathrm{GRB}}=60$ for the GOLD sample; % at $1.4\le z\le5.6$;
and $N_{\mathrm{GRB}}=78$ for the FULL sample, %  at $1.4\le z\le8.2$,
respectively.
%Unlike type Ia supernovae,
\emph{Fermi} samples provides more than 30 data points at $1.4 < z < 2.26$, whereas there is only 8 points available from the Pantheon+ sample in the same redshift interval. This feature makes GRBs particularly valuable for filling the redshift gap and for strengthening DDR test at high redshifts $z\sim8$.

\subsection{DESI DR2 BAO data}
DESI collaboration has recently released high-precision BAO measurements, providing an independent and reliable approach to trace the expansion history of the Universe. In DR2 \citep{DESI2025a,DESI2025b}, the collaboration reports constraints on the dimensionless parameter $D_{\mathrm{M}}/r_{\mathrm{d}}$, where $D_{\mathrm{M}}$ denotes the
%comoving angular diameter distance
transverse comoving distance
and $r_{\mathrm{d}}$ represents the sound horizon at the baryon-drag epoch.
The corresponding angular diameter distance can be obtained through
$
D_{\mathrm{A}}(z) = \frac{D_{\mathrm{M}}(z)}{1+z}.
$
The DESI BAO measurements %utilized in this study
are summarized in Tab~\ref{tab:desi_bao_full}.
%We use the $D_{\mathrm{A}}$ values derived from DESI BAO observations are combined with the %Pantheon+ Type Ia supernova and GRB
%luminosity distance data to test the validity of DDR.
In this work, we use the six tracers with available $D_{\rm M}/r_{\rm d}$ measurements (LRG1, LRG2, LRG3+ELG1, ELG2, QSO, and Lya) at
$0.51\le z\le2.33$
for the DDR test combined with the luminosity distance data.
%During the statistical analysis, the sound horizon scale $r_{\mathrm{d}}$ is treated as a nuisance parameter and marginalized over.

\begin{table}[htbp]
	\centering
	\footnotesize
	\captionsetup{singlelinecheck=off, justification=raggedright}
	\caption{\label{tab:desi_bao_full} DESI DR2 BAO measurements \citep{DESI2025a,DESI2025b}, including the volume-averaged distance $D_{\rm V}/r_{\rm d}$, the
		transverse comoving distance
		$D_{\rm M}/r_{\rm d}$, and the Hubble distance $D_{\rm H}/r_{\rm d}$ at various effective redshifts $z_{\rm eff}$. %In this study, only the six tracers with available $D_{\rm M}/r_{\rm d}$ measurements (LRG1, LRG2, LRG3+ELG1, ELG2, QSO, and Lya) are used for the DDR test.
	}
	\renewcommand{\arraystretch}{1.4}
	\setlength{\tabcolsep}{12pt}
	\begin{tabular}{lcccc}
		\hline\hline
		Tracer & $z_{\mathrm{eff}}$ & $D_{\mathrm{V}}/r_{\mathrm{d}}$ & $D_{\mathrm{M}}/r_{\mathrm{d}}$ & $D_{\mathrm{H}}/r_{\mathrm{d}}$ \\
		\hline
		BGS       & 0.295 & $7.942 \pm 0.075$ & -- & -- \\
		LRG1      & 0.510 & $12.720 \pm 0.099$ & $13.588 \pm 0.167$ & $21.863 \pm 0.425$ \\
		LRG2      & 0.706 & $16.050 \pm 0.110$ & $17.351 \pm 0.177$ & $19.455 \pm 0.330$ \\
		LRG3+ELG1 & 0.934 & $19.721 \pm 0.091$ & $21.576 \pm 0.152$ & $17.641 \pm 0.193$ \\
		ELG2      & 1.321 & $24.252 \pm 0.174$ & $27.601 \pm 0.318$ & $14.176 \pm 0.221$ \\
		QSO       & 1.484 & $26.055 \pm 0.398$ & $30.512 \pm 0.760$ & $12.817 \pm 0.516$ \\
		Lya       & 2.330 & $31.267 \pm 0.256$ & $38.988 \pm 0.531$ & $8.632 \pm 0.101$ \\
		\hline\hline
	\end{tabular}
	%	\begin{tablenotes}
		%		\item \small \textit{Note.}  In this study, only the six tracers with available $D_{\rm M}/r_{\rm d}$ measurements (LRG1, LRG2, LRG3+ELG1, ELG2, QSO, and Lya) are used for the DDR test.
		%	\end{tablenotes}
\end{table}

\subsection{SGL systems}
Recently, \citet{Chen2019} (hereafter C19) compiled an updated galaxy-scale sample, which contains 161 galaxy-scale SGL systems with the redshift of the source  from $0.197\le z\le3.595$, and the lens ranges from $0.064\le z\le1.004$, including 5 systems from the Lens Structure and Dynamics (LSD) survey, 26 from the Strong Lensing Legacy Survey (SL2S), 57 from the and the Sloan Lens ACS (SLACS) survey, 38 from the an extension of the SLACS survey known as “SLACS for the Masses” (S4TM), 35 from the Baryon Oscillation Spectroscopic Survey (BOSS) emission-line lens survey (BELLS).
In order to ensure the validity of the assumption of spherical symmetry on the lens galaxy, all the selected are lens galaxies in the SGL sample are ETGs with E/S0 morphologies and do not have significant substructures or close massive companion \citep{Chen2019}. %Thus, the spherically symmetric approximation is valid when modelling the lens galaxy.
More recently, \citet{Amante2020} (hereafter A20) compiled a total of 204 systems with $0.0625\le z\le0.958$ for the lens and $0.196\le z\le3.595$ for the
source, which considered 19 SLS from the CfA-Arizona Space Telescope LEns Survey (CASTLES), 107 from SLACS, 38 from BELLS, 4 from LSD, 35 from SL2S and one from the Strong-lensing Insights into Dark Energy Survey (STRIDES). %The authors used spectroscopy to select those lenses with lenticular (S0) or elliptical (E) morphologies which have been modeled assuming a SIS or SIE lens model. In their work \citep{Amante2020}, the aperture-corrected central velocity dispersions $\sigma_0$ have already been calculated.
It should be noted that SGL systems with $D_{\rm obs} > 1$ maybe unphysical \citep{Leaf2018,Amante2020,Holanda2022},
therefore, we also consider the subsample of SGL systems satisfying $D_{\rm obs} < 1$ from C19 and A20, labeled C19$^*$ including 138 systems %with $D_{\rm obs} < 1$ from C19,
and A20$^*$ including 172 systems%with $D_{\rm obs} < 1$ from C19
, respectively.
In this work, we use C19* and A20* SGL samples to test DDR.

In the singular isothermal sphere (SIS) model, the ADD ratio %with the source and the lens
can be expressed as \citep{Chen2019}
\begin{equation}
	\frac{D_{ls}}{D_s} = \frac{\theta_E \, c^2}{4 \pi \, \sigma_0^2},
\end{equation}
where $D_{ls}$ is the ADD between the lens and the source, and $D_s$ is the ADD from the observer to the source, $\theta_E$ is the Einstein radius and $\sigma_0$ (i.e, $\sigma_{e2}$) is the velocity dispersion corrected to a standard aperture (half of the effective radius), %and $c$ is the speed of light.
which is related to the observed velocity dispersion $\sigma_{\rm ap}$ by
$
\sigma_{e2} = \sigma_{\rm ap} \left( \frac{\theta_{\rm eff}}{2 \, \theta_{\rm ap}} \right)^{\eta},
$
where $\theta_{\rm eff}$ is the angular effective radius of the lens galaxy, $\theta_{\rm ap}$ is the angular size of the observational aperture, and $\eta$ is the aperture correction index. The total uncertainty of $\sigma_{e2}$ is given by
$
\sigma_{e2,{\rm tot}} = \sqrt{
	\sigma_{e2,{\rm stat}}^2 + \sigma_{e2,{\rm AC}}^2 + \sigma_{e2,{\rm sys}}^2
},
$
where $\sigma_{e2,{\rm stat}}$ denotes the statistical error propagated from the measurement uncertainty of $\sigma_{\rm ap}$, $\sigma_{e2,{\rm AC}}$ represents the error arising from the uncertainty of the aperture correction index $\eta$, and $\sigma_{e2,{\rm sys}}$ is a systematic error of approximately $3\%$, accounting for additional mass along the line of sight.

\subsection{SNe Ia Sample}
The Pantheon+ compilation provides the most up-to-date collection of SNe Ia utilized for cosmological distance determinations, which contains a total of 1701 light curves corresponding to 1550 SNe Ia within the redshift interval $0.001 < z < 2.26$ \citep{Scolnic2022}.%, enabling precise constraints on the expansion history of  Universe.
In order to prevent the impact by the peculiar velocities, 111 SN Ia at low redshifts ($z<0.01$) are removed \citep{Zhang_2024,Sousa-Neto2025}.
In this work, we use 1582 SNe Ia from Pantheon+ compilation within the redshift interval $0.01 < z < 1.4$.
The SNe Ia dataset offers %key observables such as
the apparent B-band magnitude $m_\mathrm{B}$ and the redshift $z$, which
is linked to the distance modulus $\mu$ by the standard expression:
\begin{equation}
	\mu = m_{\mathrm{B}} - M_{\mathrm{B}} = 5 \log_{10}\left(\frac{D_{\mathrm{L}}}{\text{Mpc}}\right) + 25,
	\label{eq:mu}
\end{equation}
where $D_\mathrm{L}$ is the dimensionless luminosity distance.%:
%the $M_{\mathrm{B}}$ represents the absolute magnitude of Type Ia supernovae. %This relation can be rearranged as
%$D_{\mathrm{L}} = 10^{\frac{\mu - 25}{5}}\text{Mpc}.$

\section{Reconstruction from SN Ia and GRBs}\label{sec:recon}

We use ANNs for the reconstruction task from SN Ia and GRBs to match redshifts with SGLs and BAOs. The main advantages of using ANNs are the fully data-driven nature and independence from kernel function selection. Compared with GP methods, ANN does not rely on an explicit covariance kernel.%, thereby avoiding potential biases introduced by kernel assumptions.
%\subsection{Artificial Neural Networks}
We employ an ANN framework for reconstruction from SN Ia and GRBs,
with a combined $\chi^2$ and KL divergence loss to reconstruct the apparent magnitude $m(z)$ or distance modulus $\mu(z)$ along with their uncertainties $\sigma$ \citep{Huang2025b}.
The total loss function is given by
\begin{equation}
	\mathcal{L}_{\mathrm{total}} =  \chi^{2} +  D_{\mathrm{KL}}.
\end{equation}
The $\chi^2$ term quantifies the deviation between the predicted and observed data: %defined as follows
$
\chi^2 = \boldsymbol{\Delta}^\mathrm{T} \mathbf{C}^{-1} \boldsymbol{\Delta},
\label{eq:chi2}
$
$\Delta$ is the residual vector with components: $\Delta_i = X_i - X(z_i)$, $X_i$ is the observed value, $X(z_i)$ is the theoretical value.
The KL divergence, also known as relative entropy, quantifies the difference between two probability distributions and is widely used in probabilistic modeling.
For two continuous probability densities $p(x)$ and $q(x)$, representing the observed data distribution and the model-predicted distribution respectively, the KL divergence is defined as
\begin{equation}
	\mathrm{KL}\big(p(x)\,\|\,q(x)\big)
	= \int p(x) \, \log \frac{p(x)}{q(x)} \, \mathrm{d}x \, .
\end{equation}
%KL divergence is asymmetric and does not satisfy the properties of a metric (e.g., symmetry and the triangle inequality), so it is not a true distance measure.
%Moreover, forward KL divergence, $\mathrm{KL}(p\|q)$, and reverse KL divergence, $\mathrm{KL}(q\|p)$, exhibit distinct behaviors: forward KL tends to cover all regions where $p(x)$ is non-zero, including low-probability tails, whereas reverse KL concentrates on the high-probability regions of $p(x)$, producing more focused predictions.
%These fundamental properties of KL divergence are described in \citep{Blei2017}.

%\textbf{We tailor the divergence objective to the statistical characteristics of each dataset. }
For the dense and high-precision Pantheon+ SNe, we adopt the reverse KL divergence, $\mathrm{KL}(q\|p)$.
Similar to its application in variational inference where it prioritizes precision, % \citep{Yao2025},
the reverse KL exhibits `mode-seeking' (or zero-forcing) behavior. %\citep{Minka2005}.
This encourages the model to concentrate probability mass on the dominant expansion mode, producing a sharp and focused reconstruction.
Conversely, for the sparse and scattered GRBs, we adopt the forward KL divergence, $\mathrm{KL}(p\|q)$.
Known for its `mass-covering' property, %\citep{Blei2017, Yao2025},
the forward KL forces the model distribution to encompass the entire observational support. This prevents overfitting to local clusters and ensures a robust representation of the intrinsic uncertainties, effectively capturing the large dispersion in GRB data.

Our networks employs the SiLU (Sigmoid Linear Unit) activation function \citep{Elfwing2017}, %also known as the swish function. SiLU
which is smooth and non-monotonic to ensure stable gradient flow during training and reduces oscillations in the predictions:
$
\text{SiLU}(x) = x \cdot \sigma(x) = \frac{x}{1 + e^{-x}},
$
where $\sigma(x)$ is the standard sigmoid function.

%In order to select the optimal network architectures, we evaluate multiple candidate neural network architectures using the full Pantheon+ SNe Ia sample and the GOLD sample as the test set.
In order to select the optimal network architectures, we evaluate multiple candidate neural network architectures using the full Pantheon+ SNe Ia sample and the GRB sample.
During training, both the training and validation losses are monitored, and an early stopping criterion is applied to prevent overfitting and ensure good generalization.
Since the $\chi^2$ loss depends on the full Pantheon+ covariance matrix, which has been verified to be positive definite and remains unchanged throughout training, we invert the covariance matrix only once. During training, we employ full-batch gradient descent for each epoch. This approach not only ensures that the loss function accurately captures the correlations between data points, but also significantly reduces computational cost.
For the Pantheon+ data, the optimal neural network has four fully connected layers with sizes [128, 128, 64, 32],
while for the GRB data, the optimal network consists of three fully connected layers with sizes [128, 128, 128].
These architectures achieve lower losses %on their respective datasets
without any signs of overfitting.
%We give the details for the selection of the ANN architecture, the training-validation process with mock data in Appendix A.
We give the details for the selection of the optimal ANN architecture, with mock data used to test its performance in Appendix A.

\subsection{Reconstruction of $m(z)$ from SNe Ia}

We reconstruct the apparent magnitude $m(z)$ or distance modulus $\mu(z)$ from the Pantheon+ observational data using the
% artificial neural network
ANN
developed in this work. The network is trained with a combined $\chi^2$ and KL divergence loss to simultaneously obtain the reconstructed $m(z)$ or $\mu(z)$ and their uncertainties $\sigma$.
In the low-redshift range ($z \le 1.4$), we use the Pantheon+ supernova sample.
The reconstructed apparent magnitude $m(z)$ from SNe Ia at $0.01< z \le 1.4$ is shown in Fig~\ref{fig:sn_grb}.
We also show the reconstructed results from SNe at $0.01< z \le 2.28$ to the maximum redshift for SGL systems $z=3.6$.
For each supernova, the observed distance modulus is given by
\begin{equation}
	\mu_i = m_{{\rm B},i}^{\mathrm{corr}} - M_{\rm B},
\end{equation}
%where $m_{{\rm B},i}^{\mathrm{corr}}$ is the corrected apparent magnitude, and $M_{\rm B}$ is taken from two independent studies that constrained $M_{\rm B}$ using the Pantheon+ sample \citep{Mukherjee_2024,Riess2022}.
where $m_{{\rm B},i}^{\mathrm{corr}}$ is the corrected apparent magnitude.
The error in the distance modulus is calculated as:
$
\sigma_\mu = \sqrt{\sigma_{m_{\rm B}}^2 + \sigma_{M_{\rm B}}^2},
$
where \(\sigma_{m_{\rm B}}\) and \(\sigma_{M_{\rm B}}\) are the uncertainties in the apparent and absolute magnitudes, respectively.
The absolute magnitude can be calibrated by setting an absolute distance scale with primary distance anchors such as Cepheids.
%The SH0ES (Supernovae and $H_0$ for the Equation of State of dark energy) team obtained $M_B = -19.253\pm0.027$ \citep{Riess2022} from the Pantheon+ sample \citep{Scolnic2022}.
We consider the values of $M_{\rm B}$ %are adopted
from two independent studies based on the Pantheon+ sample: $M_{\rm B}^{\rm SH0ES} =  -19.253 \pm 0.027~\mathrm{mag}$ from SH0ES (Supernovae and $H_0$ for the Equation of State of dark energy) \citep{Riess2022}, and $M_{\rm B}^{\rm M24} = -19.353^{+0.073}_{-0.078}~\mathrm{mag}$ from from the Pantheon+ and OHD \citep{Mukherjee_2024}(Hereafter, we refer to the calibration simply as M24).
%The
%For the ANN training of reconstruction from SNe Ia, we distinguish two cases:
%i) Training $m(z)$: only the observational covariance matrix  $C_{\mathrm{tot}} = C_{\mathrm{obs}}$ (i.e., the original Pantheon+ covariance matrix) is used to learn the variation of the apparent magnitude with redshift.
%Figure~\ref{fig:m_recon} shows the reconstructed apparent magnitude
%$m(z)$ of the Pantheon+ sample.
%{ii) Training $\mu(z)$: To test the DDR when $M_{\rm B}$ fixed, and to take into account the correlations between the observational data (i.e., the covariance matrix), we choose to reconstruct $\mu(z)$ directly rather than substituting $M_{\rm B}$ into the $\chi^2$ function used for testing the DDR. The total covariance matrix is given by
	%\begin{equation}
	%	C_{\rm tot} = C_{\rm obs} + \sigma_{M_{\rm B}}^2 I,
	%\end{equation}
	%where $\sigma_{M_{\rm B}}$ is the uncertainty of the SN absolute magnitude, and $I$ is the identity matrix.

	\subsection{Reconstruction of $\mu(z)$ from GRBs}
	For the high-redshift range ($z > 1.4$), we reconstruct the distance modulus $\mu(z)$ using only GRB data.
	Since the GRB data points are independent, %and no covariance matrix is provided,
	only the KL divergence loss is used to train the ANN.
	%ANN.
	We use \emph{Fermi} FULL and GOLD samples %are trained separately
	to obtain $\mu(z)$ and its uncertainties in the high-redshift regime.
	The reconstructed distance modulus for FULL and GOLD samples at $z > 1.4$ is shown in Fig~\ref{fig:sn_grb}.
	
%	\begin{figure*}
%		\centering
%		% 第一列：m(z)
%		\begin{minipage}{0.32\textwidth}
%			\centering
%			\includegraphics[width=\textwidth]{./m_recon.png}
%			
%		\end{minipage}
%		\hfill
%		% 第二列：GRB FULL
%		\begin{minipage}{0.32\textwidth}
%			\centering
%			\includegraphics[width=\textwidth]{./FULL.png}
%			
%		\end{minipage}
%		\hfill
%		% 第三列：GRB GOLD
%		\begin{minipage}{0.32\textwidth}
%			\centering
%			\includegraphics[width=\textwidth]{./GOLD.png}
%			
%		\end{minipage}
%		
%		\captionsetup{singlelinecheck=off, justification=raggedright}
%		\caption{ANN reconstructions: \emph{left} panel for $m(z)$ from Pantheon+ at $0.01< z \le 1.4$; \emph{middle} and \emph{right} panels for GRB distance modulus $\mu(z)$ from \emph{Fermi} FULL and GOLD samples at $z > 1.4$, respectively.}
%		\label{fig:sn_grb}
%		
%		
%	\end{figure*}

\begin{figure*}
	\centering
	% 第一行：m(z) 和 GRB FULL
	\begin{minipage}{0.48\textwidth}
		\centering
		\includegraphics[width=\textwidth]{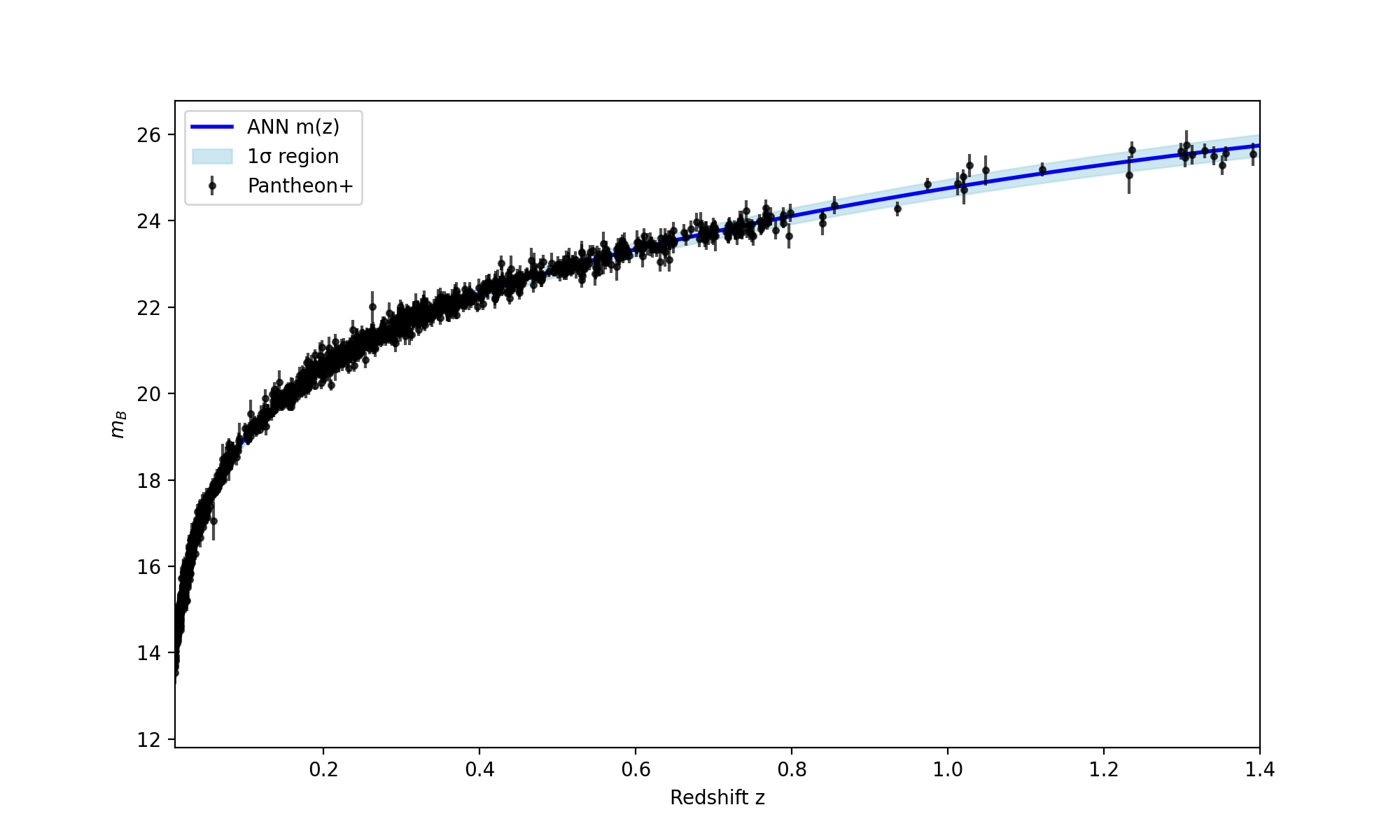}
		\subcaption{SN(0.01<z<1.4)}
	\end{minipage}
	\hfill
	\begin{minipage}{0.48\textwidth}
		\centering
		\includegraphics[width=\textwidth]{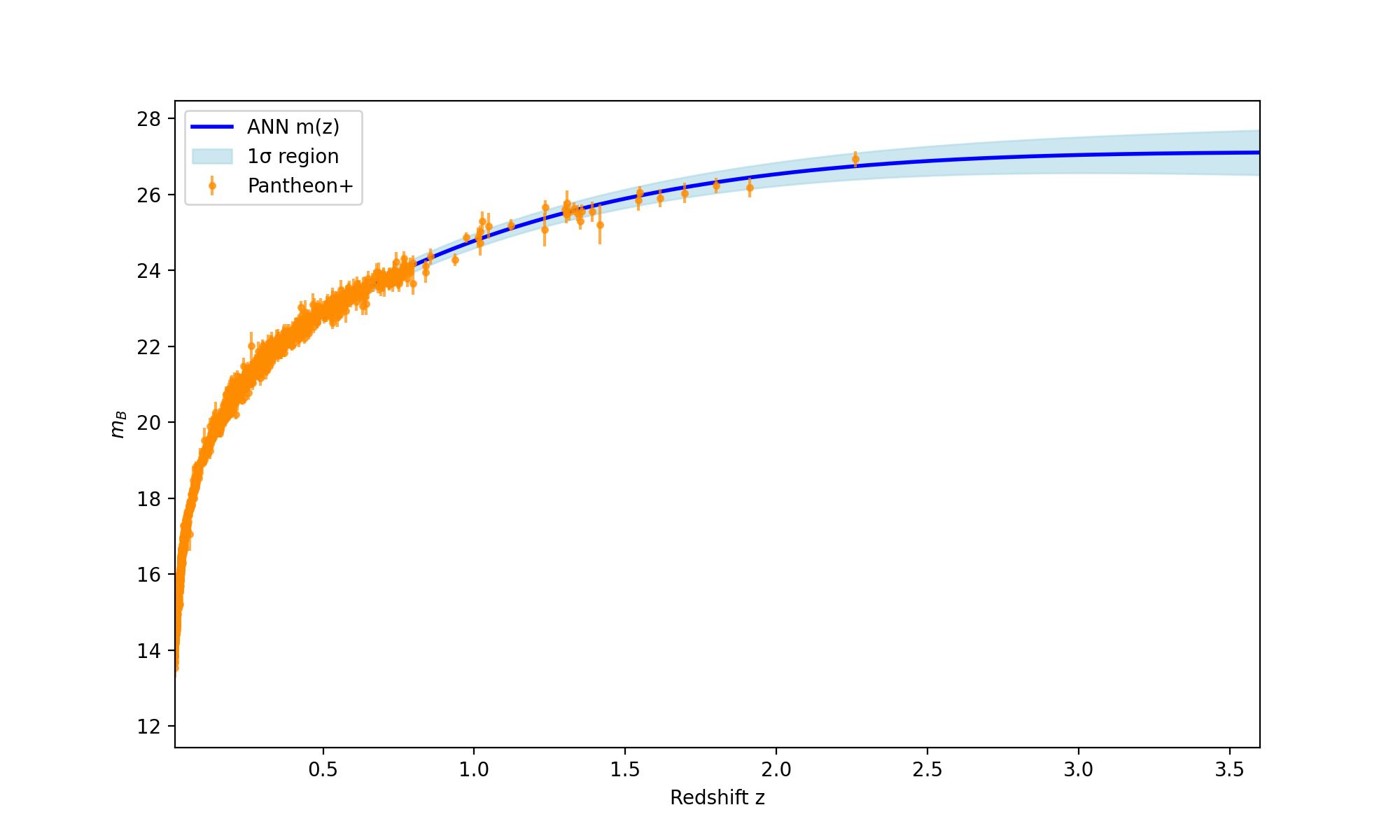}
		\subcaption{SN(0.01<z<3.6)}
	\end{minipage}
	
	% 第二行：GRB GOLD 和预留位置
	\begin{minipage}{0.48\textwidth}
		\centering
		\includegraphics[width=\textwidth]{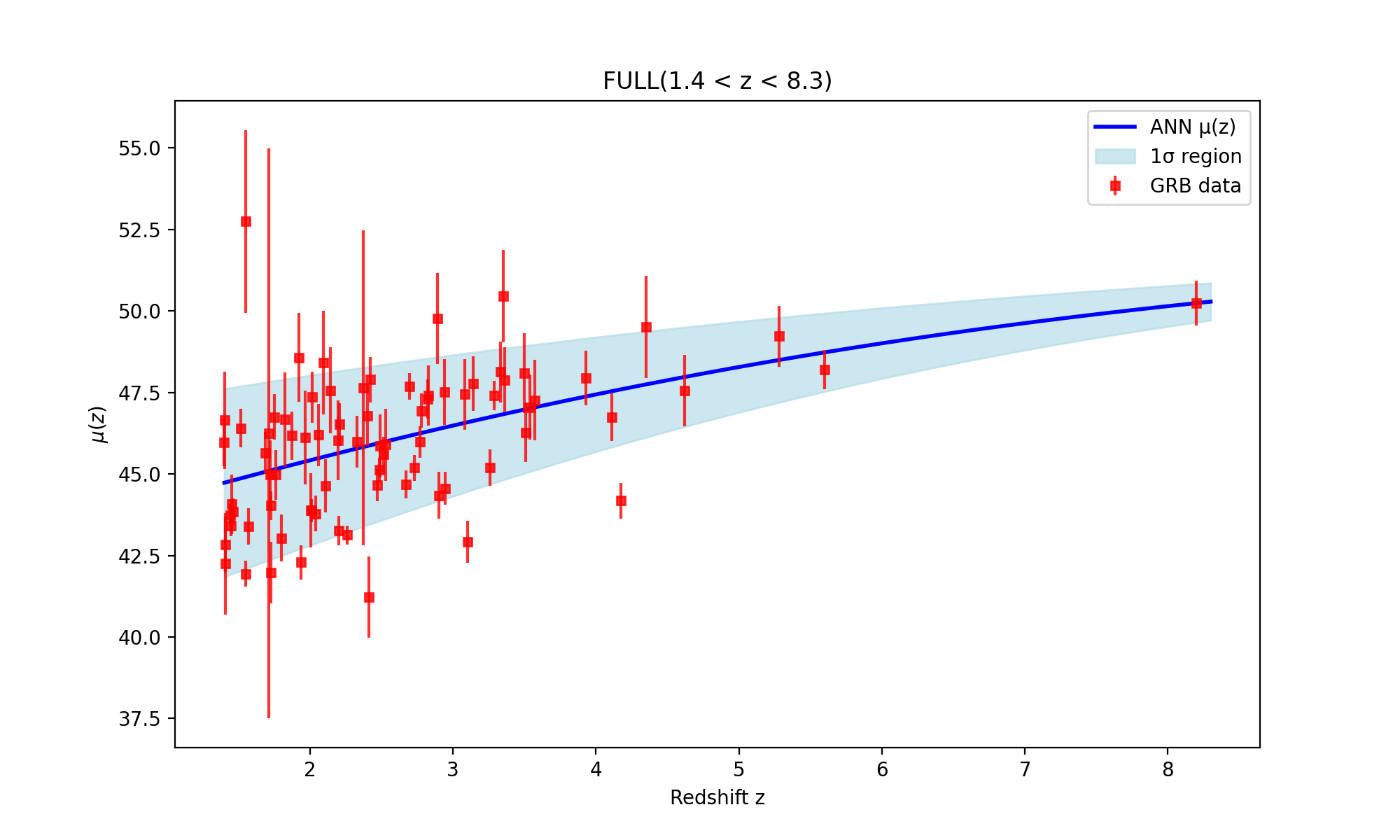}
		\subcaption{FULL(1.4<z<8.3)}
	\end{minipage}
	\hfill
	\begin{minipage}{0.48\textwidth}
		\centering
		% 预留位置用于插入新图片
		 \includegraphics[width=\textwidth]{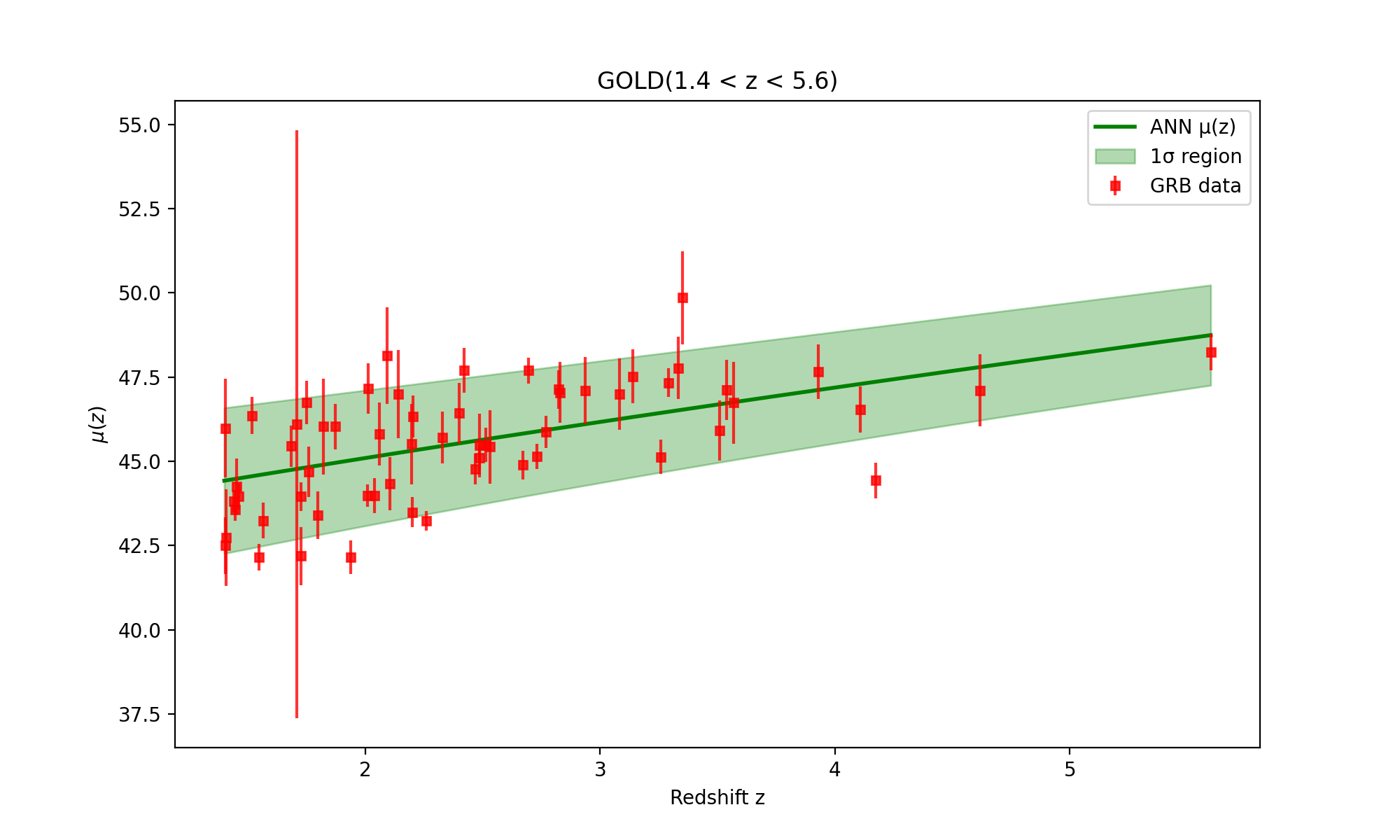}
		 \subcaption{GOLD(1.4<z<5.6)}
	\end{minipage}
	
	\captionsetup{singlelinecheck=off, justification=raggedright}
%	\caption{ANN reconstructions: \emph{left} panel for $m(z)$ from Pantheon+ at $0.01< z \le 1.4$; \emph{middle} and \emph{right} panels for GRB distance modulus $\mu(z)$ from \emph{Fermi} FULL and GOLD samples at $z > 1.4$, respectively.}
     \caption{ANN reconstructions from Pantheon+ and \emph{Fermi} samples. \emph{Upper Panels} show reconstructions of $m(z)$  from Pantheon+ at $0.01 < z \le 1.4$ and $0.01 < z \le 3.6$; \emph{Lower Panels} show reconstructions of distance modulus $\mu(z)$ from the \emph{Fermi} FULL and GOLD samples at $z > 1.4$, respectively.}
\label{fig:sn_grb}
	\label{fig:sn_grb}
\end{figure*}
%the $M_{\mathrm{B}}$ represents the absolute magnitude of Type Ia supernovae. %This relation can be rearranged as
%$D_{\mathrm{L}} = 10^{\frac{\mu - 25}{5}}\text{Mpc}.$

\section{Testing the DDR}\label{sec:test_ddr}
In the work, we adopt the parameterizations of $\eta$ values to test DDR. %In these models, the standard DDR corresponds to $\eta_0 \equiv 0$, and any significant deviation from this baseline could indicate the presence of exotic physics.
%\subsection{Parameterizations of CDDR}
We compare the observed $\eta$ values with two commonly parameterizations, the predictions are denoted as P1 model ($\eta(z) = 1 + \eta_0 z$) \citep{Bassett2004} and P2 model ($\eta(z) = 1 + \eta_0 z/(1+z)$) \citep{Holanda2010}
%\subsection{Testing the DDR with parameterizations}
%P1: $\eta(z) = 1 + \eta_0 z$; \\
%P2: $\eta(z) = 1 + \eta_0 z/(1+z)$.
%P4: $\eta(z) = (1+z)^{\eta_0}$ \citep{Alfano2025b}.
%In this work, we compute $\eta(z)$ using
with observational data from SN Ia, GRBs, BAOs, and SGLs.
For SN Ia, the luminosity distance $D_{\rm L}(z)$ is derived from the apparent magnitude $m(z)$ and the absolute magnitude $M_{\rm B}$,
while for GRBs, $D_{\rm L}(z)$ is obtained from the distance modulus $\mu(z)$. The angular diameter distance $D_{\rm A}$ is calculated from
the transverse comoving distance $D_{\rm M}(z)/r_{\rm d}$ obtained from BAO measurements.
%with the sound horizon fixed at
%$r_{\rm d} = 147.78~{\rm Mpc}$ \citep{Planck2018VI}.
The observed value of $\eta$ is given by
\begin{equation}
	\eta_{\rm obs}(z) = \frac{D_{\rm L}(z)}{D_{\rm A}(z) (1+z)^2},
\end{equation}

The uncertainty of the observed $\eta$ values is calculated from the relative errors of the luminosity distance $D_{\rm L}(z)$ and angular diameter distance $D_{\rm A}(z)$ as
\begin{equation}
	\sigma_{\eta_{\rm obs}}^2 = \eta_{\rm obs}^2 \left[
	\left( \frac{\sigma_{D_{\rm L}}(z)}{D_{\rm L}(z)} \right)^2 +
	\left( \frac{\sigma_{D_{\rm A}}(z)}{D_{\rm A}(z)} \right)^2
	\right],
\end{equation}
where $\sigma_{D_{\rm L}}$ and $\sigma_{D_{\rm A}}$ denote the observational uncertainties of $D_{\rm L}$ and $D_{\rm A}$, respectively.

For the SGL systems, the observed quantity is:
$
D_{\rm obs} \equiv \frac{D_{ls}}{D_s},
$
and the model distance is given by \cite{Holanda2022}
% \citep{Liao2016,Fu2017}
\begin{equation}
	D_{\rm model} = 1 - \frac{D_L^l}{D_L^s} \cdot \frac{1+z_s}{1+z_l} \cdot \frac{\eta_{\rm th}(z_s)}{\eta_{\rm th}(z_l)},
\end{equation}
where $D_L^l$ and $D_L^s$ are the luminosity distances at the lens and source redshifts, $z_l$ and $z_s$, respectively, which are reconstructed from SN and GRB observations at these redshifts, $\eta_{\rm th}(z)$ are parameterizations of DDR with the P1 or P2 model.
The $\chi^2$ statistic using SGLs with SNe Ia+GRBs is constructed as
\begin{equation}
	\chi^2_{\rm SGL} =
	\sum_{\rm SGL} \frac{(D_{\rm obs} - D_{\rm model})^2}{D_{\rm err}^2 + \sigma_{\rm model}^2}
\end{equation}
Here $D_{\rm err}$ denotes the observational uncertainty of $D_{\rm obs}$, and $\sigma_{\rm model}$ denotes the propagated uncertainty of $D_{\rm model}$.
The $D_{\rm err}$ is given by \citep{Amante2020}:
\begin{equation}
	D_{\rm err} = D_{\rm obs} \sqrt{ \left( \frac{\sigma_{\theta_{\rm E}}}{\theta_{\rm E}} \right)^2 + 4 \left( \frac{\sigma_{\sigma_{0}}}{\sigma_{0}} \right)^2 },
\end{equation}
where $\sigma_{\theta_{\rm E}}$ is the uncertainty of the Einstein radius, fixed at a fractional level of 5\%
\citep{Cao2015,Liao2016,Holanda2022}.
The uncertainty $\sigma_{\rm model}$ is derived via error propagation with respect to the reconstructed luminosity distances $D_{L}^{l}$ and $D_{L}^{s}$:
\begin{equation}
	\begin{aligned}
		\sigma_{\rm model} &= \sqrt{ \left( \frac{\partial D_{\rm model}}{\partial D_{L}^{l}} \sigma_{D_{L}^{l}} \right)^2 + \left( \frac{\partial D_{\rm model}}{\partial D_{L}^{s}} \sigma_{D_{L}^{s}} \right)^2 } \\
		&= \sqrt{ \left( - \frac{1}{D_{L}^{s}} \left[ \frac{1+z_s}{1+z_l} \frac{\eta_{th}(z_s)}{\eta_{th}(z_l)} \right] \sigma_{D_{L}^{l}} \right)^2 + \left( \frac{D_{L}^{l}}{(D_{L}^{s})^2} \left[ \frac{1+z_s}{1+z_l} \frac{\eta_{th}(z_s)}{\eta_{th}(z_l)} \right] \sigma_{D_{L}^{s}} \right)^2 },
	\end{aligned}
\end{equation}
where $\sigma_{D_{L}^{l}}$ and $\sigma_{D_{L}^{s}}$ are the uncertainties of the luminosity distances from the ANN at the lens and source redshifts, respectively.
%For SGL systems with $D_{\rm obs} > 1$ maybe unphysical \citep{Leaf2018,Amante2020,Holanda2022},
%the datasets labeled A20 and A20$^*$ correspond to the full sample and the subsample of 172 systems satisfying $D_{\rm obs} < 1$ from A20, respectively. Similarly, C19 and C19$^*$ refer to the full sample and the subset of 138 systems with $D_{\rm obs} < 1$ from C19, respectively.
%Figures~\ref{fig:three}--\ref{fig:one} present the results for the FULL and GOLD samples under the three schemes.
%Specifically, the datasets labeled 204-SGL and 172-SGL correspond to the full sample and the subsample satisfying $D_{\rm obs} < 1$ from A20, respectively. Similarly, 161-SGL and 138-SGL refer to the full sample and the $D_{\rm obs} < 1$ subset from C19, respectively.
%The $\chi^2$ statistic using BAOs with SNe Ia+GRBs is constructed as
%\begin{equation}
%	\chi^2_{\rm BAO} =
%	\sum_{\rm BAO} \frac{(\eta_{\rm obs}(z) - \eta_{\rm th}(z))^2}{\sigma_{\eta_{\rm obs}}^2},
%\end{equation}
The total $\chi^2$ statistic using SGLs+BAOs with SNe Ia+GRBs is constructed as
\begin{equation}
	\chi^2_{\rm total} = \chi^2_{\rm SGL} +\sum_{\rm BAO} \frac{(\eta_{\rm obs}(z) - \eta_{\rm th}(z))^2}{\sigma_{\eta_{\rm obs}}^2}.
\end{equation}

\begin{table*}
	\centering
	\tiny
	\captionsetup{singlelinecheck=off, justification=raggedright}
	\caption{\label{table-sorted-clean}Fitting results of the $P_1$ and $P_2$ models using SGLs+BAOs with SNe Ia+GRBs ($1\sigma$ and $2\sigma$ confidence levels) %. The results are ordered
		by free $M_{\rm B}$, and $M_{\rm B}$ fixed to M24 or SH0ES, respectively.}
	
	\renewcommand{\arraystretch}{1.8}
	\setlength{\tabcolsep}{4pt}
	
	\resizebox{\textwidth}{!}{%
		\begin{tabular}{l cccc}
			\hline
			\hline
			\multirow{2}{*}{Dataset Combination}
			& \multicolumn{2}{c}{$P_1$ model} & \multicolumn{2}{c}{$P_2$ model} \\
			\cmidrule(lr){2-3} \cmidrule(lr){4-5}
			& $\eta_0$ & $M_{\rm B}$ & $\eta_0$ & $M_{\rm B}$ \\
			\hline
			
			SN + FULL + A20$^*$ + BAO
			& $-0.05^{+0.03+0.07}_{-0.03-0.06}$ & $-19.38^{+0.09+0.19}_{-0.10-0.20}$
			& $-0.19^{+0.09+0.20}_{-0.09-0.17}$ & $-19.28^{+0.12+0.23}_{-0.12-0.24}$ \\
			
			SN + GOLD + A20$^*$ + BAO
			& $-0.07^{+0.03+0.07}_{-0.03-0.05}$ & $-19.33^{+0.09+0.18}_{-0.10-0.20}$
			& $-0.24^{+0.09+0.19}_{-0.08-0.16}$ & $-19.22^{+0.11+0.22}_{-0.12-0.24}$ \\
			
			SN + FULL + C19$^*$ + BAO
			& $0.01^{+0.05+0.11}_{-0.05-0.09}$ & $-19.45^{+0.11+0.22}_{-0.12-0.23}$
			& $0.01^{+0.13+0.26}_{-0.12-0.23}$ & $-19.44^{+0.13+0.27}_{-0.14-0.28}$ \\

			SN + GOLD + C19$^*$ + BAO
			& $-0.02^{+0.05+0.10}_{-0.04-0.08}$ & $-19.40^{+0.11+0.21}_{-0.11-0.23}$
			& $-0.05^{+0.12+0.25}_{-0.11-0.21}$ & $-19.38^{+0.13+0.26}_{-0.13-0.27}$ \\
			
			SN($z\le3.6$)  + C19$^*$ + BAO
			
			& $0.03^{+0.03+0.07}_{-0.03-0.06}$ & $-19.48^{+0.10+0.20}_{-0.10-0.21}$
			& $0.10^{+0.12+0.24}_{-0.11-0.22}$ & $-19.52^{+0.13+0.27}_{-0.14-0.28}$ \\
			\hline
			
			SN $\times M_{\rm B}^{\rm M24}$ + FULL + A20$^*$ + BAO
			& $-0.05^{+0.03+0.07}_{-0.03-0.06}$ & --
			& $-0.16^{+0.08+0.16}_{-0.08-0.15}$ & -- \\
			
			SN $\times M_{\rm B}^{\rm M24}$ + GOLD + A20$^*$ + BAO
			& $-0.07^{+0.03+0.06}_{-0.03-0.06}$ & --
			& $-0.18^{+0.08+0.15}_{-0.08-0.15}$ & -- \\
			
			SN $\times M_{\rm B}^{\rm M24}$ + FULL + C19$^*$ + BAO
			& $-0.01^{+0.04+0.08}_{-0.04-0.07}$ & --
			& $-0.05^{+0.08+0.17}_{-0.08-0.16}$ & -- \\

			SN $\times M_{\rm B}^{\rm M24}$ + GOLD + C19$^*$ + BAO
			& $-0.03^{+0.04+0.07}_{-0.04-0.07}$ & --
			& $-0.06^{+0.08+0.17}_{-0.08-0.16}$ & -- \\
			
			SN($z\le3.6$) $\times M_{\rm B}^{\rm M24}$  + C19$^*$ + BAO
			& $0.00^{+0.02+0.05}_{-0.02-0.05}$ & --
			& $-0.01^{+0.07+0.13}_{-0.06-0.13}$ & -- \\
			\hline
			
			SN $\times M_{\rm B}^{\rm SH0ES}$ + FULL + A20$^*$ + BAO
			& $-0.06^{+0.03+0.06}_{-0.03-0.05}$ & --
			& $-0.20^{+0.07+0.14}_{-0.07-0.14}$ & -- \\
			
			SN $\times M_{\rm B}^{\rm SH0ES}$ + GOLD + A20$^*$ + BAO
			& $-0.08^{+0.03+0.06}_{-0.03-0.05}$ & --
			& $-0.22^{+0.07+0.14}_{-0.07-0.14}$ & -- \\
			
			SN $\times M_{\rm B}^{\rm SH0ES}$ + FULL + C19$^*$ + BAO
			& $-0.04^{+0.03+0.07}_{-0.03-0.06}$ & --
			& $-0.12^{+0.08+0.15}_{-0.08-0.15}$ & -- \\
			
			SN $\times M_{\rm B}^{\rm SH0ES}$ + GOLD + C19$^*$ + BAO
			& $-0.05^{+0.03+0.07}_{-0.03-0.06}$ & --
			& $-0.13^{+0.08+0.15}_{-0.07-0.15}$ & -- \\
			
%            SN($z\le3.6$) $\times M_{\rm B}^{\rm SH0ES}$ + C19$^*$ + BAO
%            & $-0.02^{+0.02+0.04}_{-0.02-0.04}$ & --
%            & $-0.08^{+0.06+0.12}_{-0.06-0.12}$ & -- \\

            SN($z\le3.6$) $\times M_{\rm B}^{\rm SH0ES}$ + C19$^*$ + BAO
            & $-0.02^{+0.02+0.04}_{-0.02-0.04}$ & --
            & $-0.07^{+0.06+0.12}_{-0.06-0.12}$ & -- \\
			\hline
			\hline
		\end{tabular}
		\label{tab:tot}
	}
\end{table*}
\begin{figure*}
	\centering
	% 单行两列
	\begin{minipage}{0.4\textwidth}
		\centering
		\includegraphics[width=\textwidth]{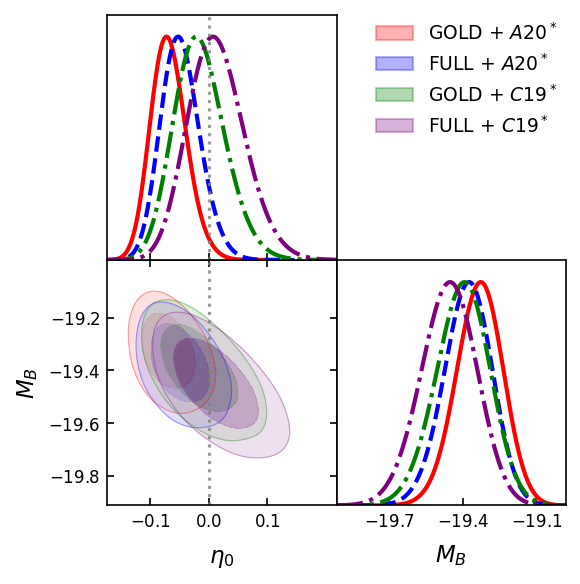}
		\subcaption{P1 model}
	\end{minipage}
	\hfill
	\begin{minipage}{0.4\textwidth}
		\centering
		\includegraphics[width=\textwidth]{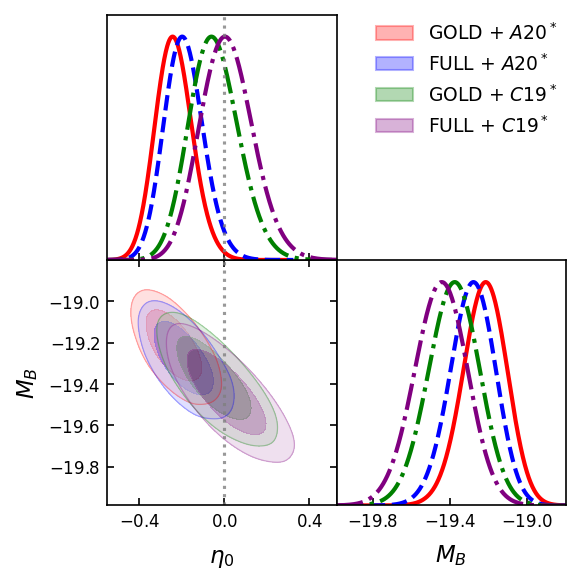}
		\subcaption{P2 model}
	\end{minipage}
	
	\captionsetup{singlelinecheck=off, justification=raggedright}
	\caption{Constraints on the cosmological parameters $\eta_0$, $M_{\rm B}$ using SGLs+BAOs with SNe Ia+GRBs. The left panel shows results for the P1 model, and the right panel shows results for the P2 model.}
	\label{fig:two}
\end{figure*}

\begin{figure*}
	\centering
	% 第一行：两列
	\begin{minipage}{0.4\textwidth}
		\centering
		\includegraphics[width=\textwidth]{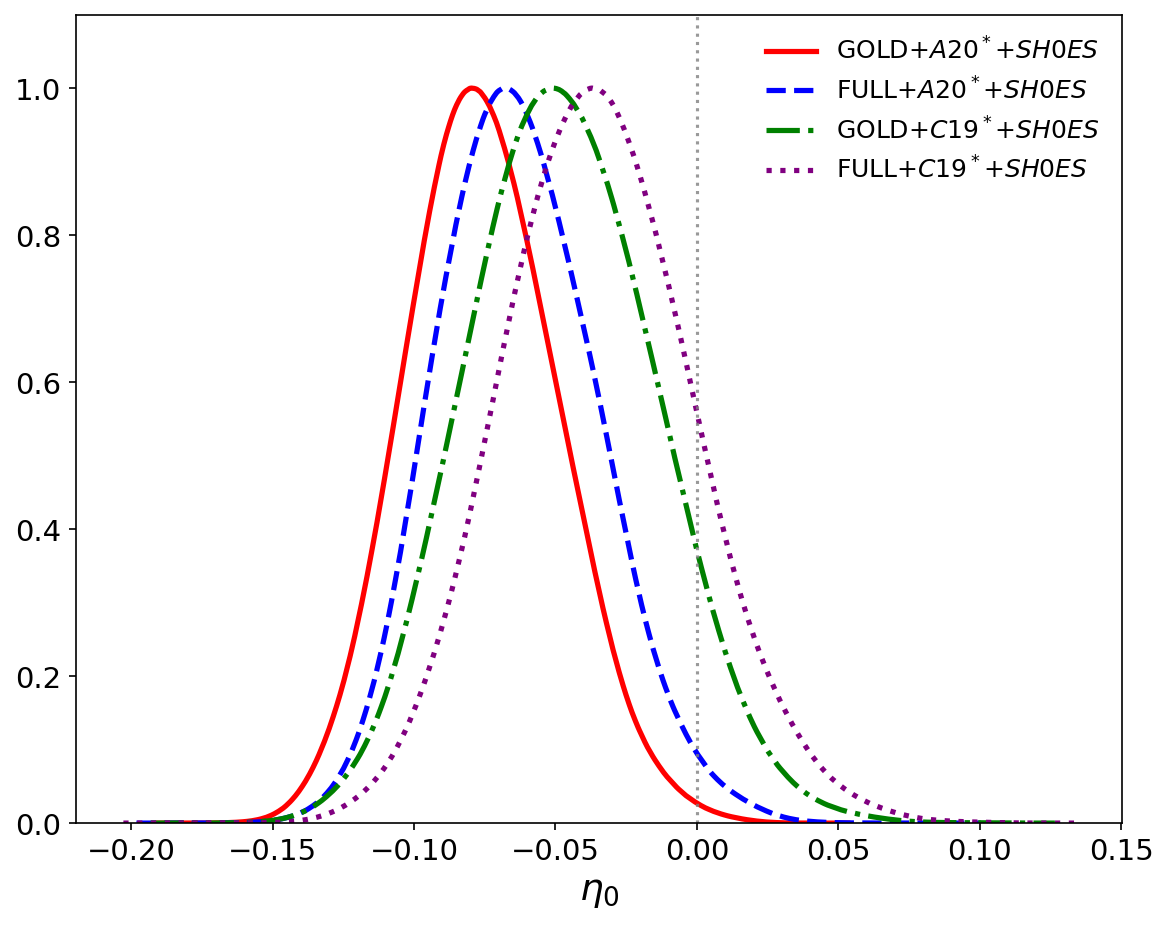}
		\subcaption{P1 model}
	\end{minipage}
	\hfill
	\begin{minipage}{0.4\textwidth}
		\centering
		\includegraphics[width=\textwidth]{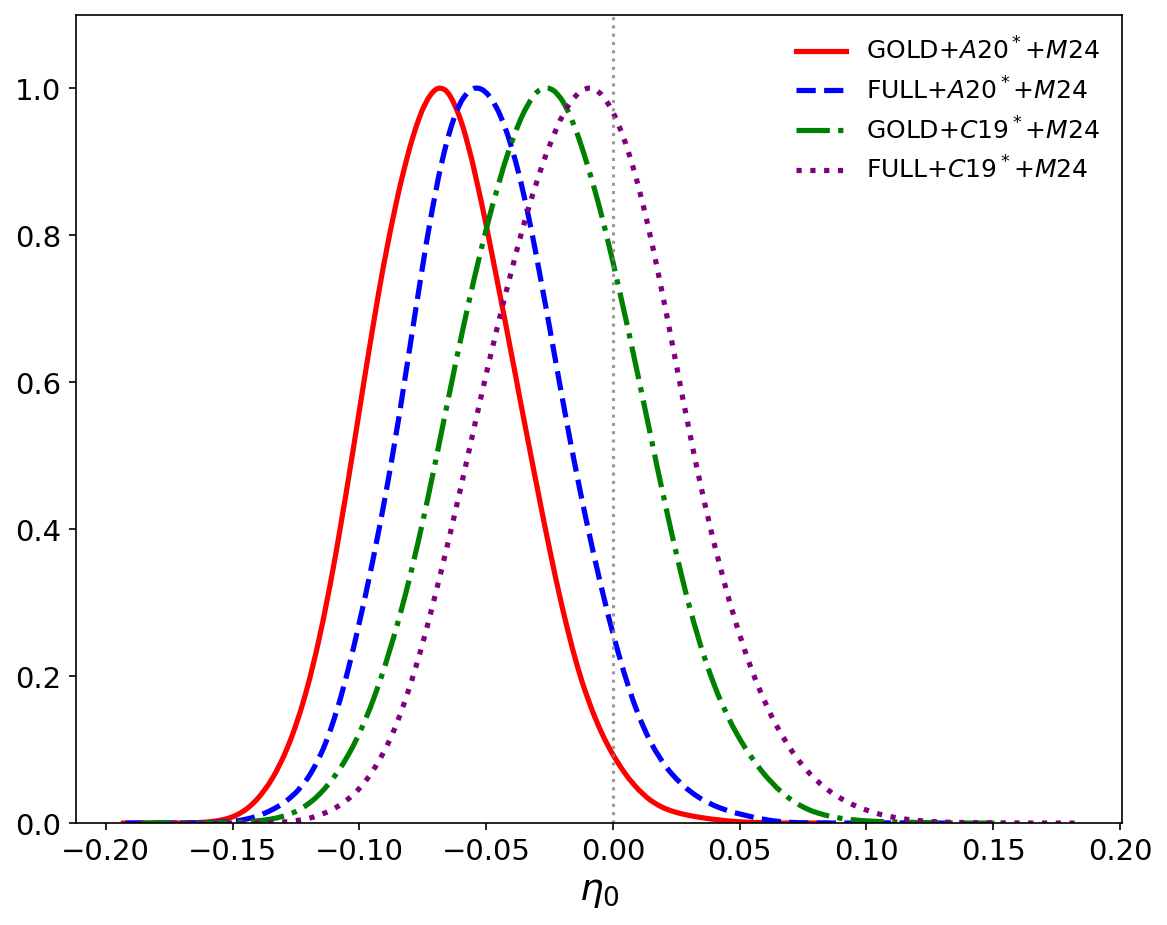}
		\subcaption{P1 model}
	\end{minipage}
	
	\vspace{0.3cm} % 行间距
	
	% 第二行：两列
	\begin{minipage}{0.4\textwidth}
		\centering
		\includegraphics[width=\textwidth]{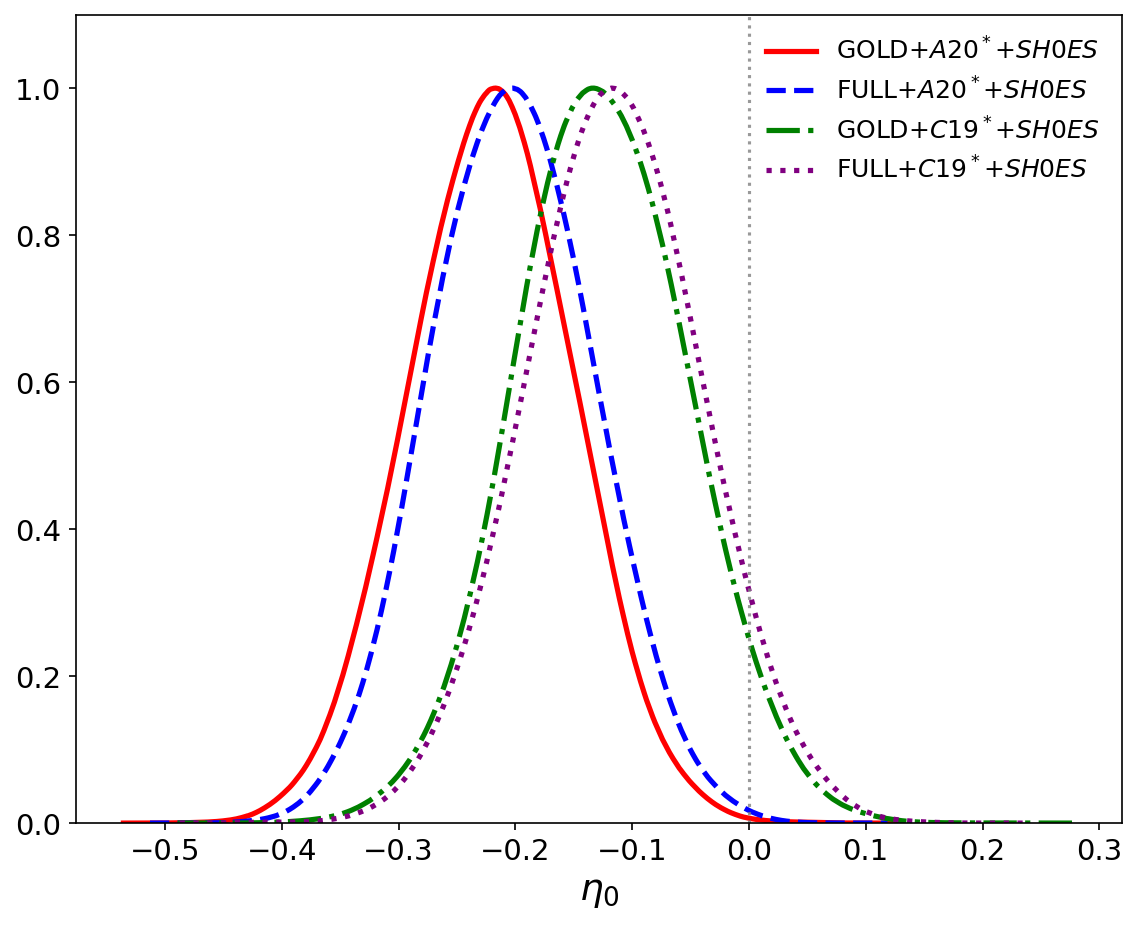}
		\subcaption{P2 model}
	\end{minipage}
	\hfill
	\begin{minipage}{0.4\textwidth}
		\centering
		\includegraphics[width=\textwidth]{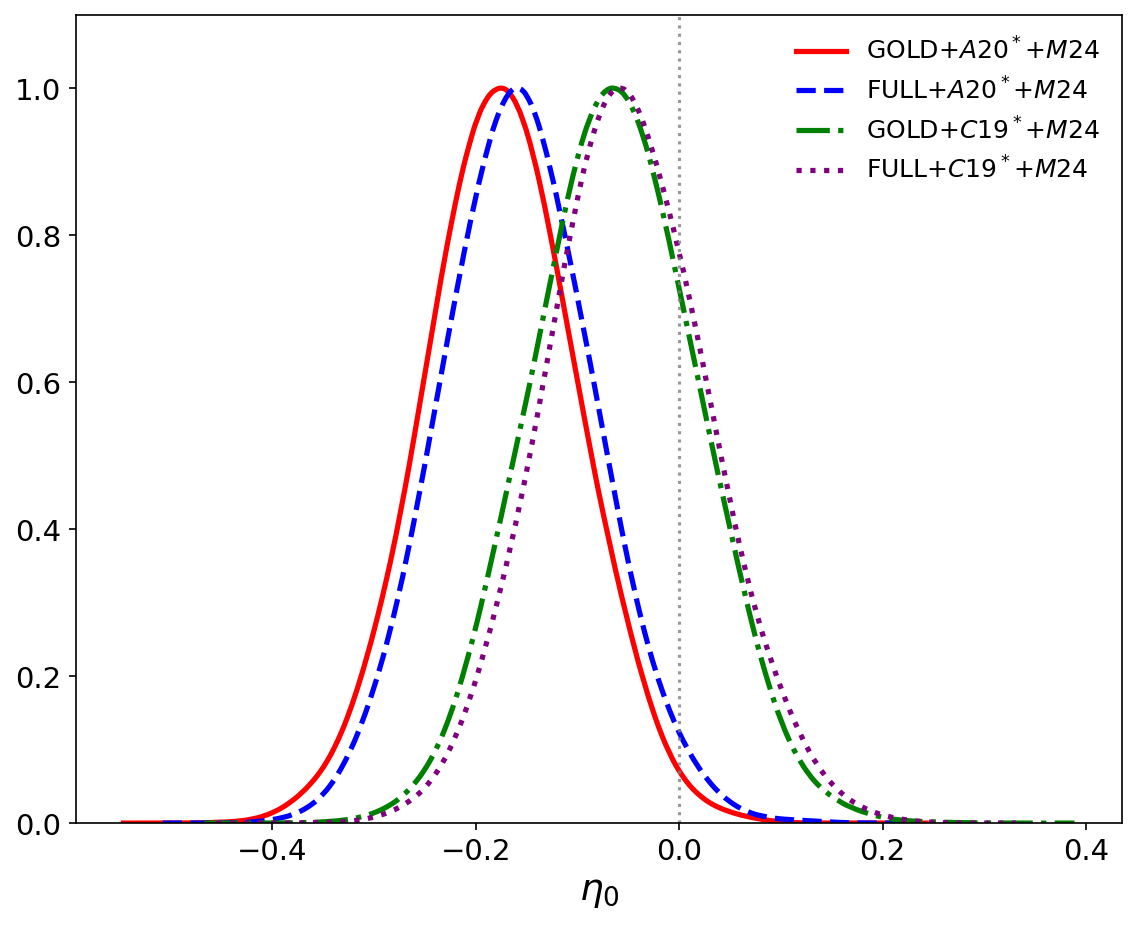}
		\subcaption{P2 model}
	\end{minipage}
	
%	\caption{
%		The 1D marginalized posterior distributions of the $\eta_0$ parameter from the $M_{\rm B}^{\rm M24}$ and $M_{\rm B}^{\rm SH0ES}$ using SGLs+BAOs with SNe Ia+GRBs.
%	}
    \captionsetup{singlelinecheck=off, justification=raggedright}
    \caption{The 1D marginalized posterior distributions of the DDR parameter $\eta_0$ from the joint analysis of SGLs+BAOs with SNe Ia+GRBs. The top and bottom rows show the results for the P1 and P2 parameterizations, respectively. For each parameterization, the left (right) panel corresponds to the analysis using the $M_{\rm B}^{\rm SH0ES}$ ($M_{\rm B}^{\rm M24}$) calibration.}

	\label{fig:one}
\end{figure*}

We employ the Markov Chain Monte Carlo (MCMC) from \cite{ForemanMackey2013} method to constrain the model parameters.
We adopt uniform priors of $M_{\rm B} \in [-21, -18]$ as a free parameter
or fix $M_{\rm B}$(M24 or SH0ES) and $r_{\rm d} = 147.09~\mathrm{Mpc}$\footnote{
\citet{Li2025} used $r_d = 147.78~\mathrm{Mpc}$ from Planck, and neglecting its uncertainty does not affect the DDR test results.
We adopt the constrained value $r_d = 147.09 \pm 0.26~\mathrm{Mpc}$ from Planck TT, TE, EE+lowE+lensing \citet{Aghanim2020}. It should be note that $r_{\rm d}$ can also be treated as a free parameter~(\citet{L2020})
and the uncertainty of $r_d$ can propagate into the DDR tests. The impact of $r_{\rm d}$ treating as a free parameter and the uncertainty of $r_d$ on the DDR tests are presented in Appendix~C.
%By comparing with the $r_d = 147.09~\mathrm{Mpc}$ result, we find that neglecting the uncertainty does not influence the test outcomes(See Tab~\ref{tab:tot}).
%$r_{\rm d}$ can also be treated as a free parameter~(\citet{L2020}), and the results for this case are shown in Fig~\ref{fig:two}.
},
% \footnote{Following
%\citet{Li2025}, we adopt $r_{\rm d} = 147.78~\mathrm{Mpc}$ from Planck and neglect its uncertainty, since it would have a negligible impact on the results of test DDR.}
while assuming a flat uninformative prior for $\eta_0$.
The fitting results using SGLs(C19*, A20*)+BAOs with SNe Ia+GRBs(FULL, GOLD)  are shown in Fig~\ref{fig:two},
Fig~\ref{fig:one} and summarized in Tab~\ref{tab:tot}.
In order to show the specific contributions of GRB data, particularly how GRBs constrain DDR at high redshifts, we also show the results from data only using  SNe reconstructed at $z\le3.6$ with C19* + BAOs.

We find that the fitting results using combinations with the FULL sample are consistent with the GOLD sample.
We also find combinations with the C19* sample slightly favoured the DDR over combinations with the A20* sample:
%When combinations with $D_{\rm obs} > 1$ are not excluded (C19*, A20*), the DDR shows significant deviations.
%When only $D_{\rm obs} < 1$ systems are retained, in the cases of free  $M_{\rm B}$ with $r_{\rm d} = 147.78~\mathrm{Mpc}$,
the C19$^*$ subsample combined with the FULL or GOLD samples
satisfies the DDR within $1\sigma$, while the A20* subsample %+ FULL satisfies it
close to $2\sigma$.
When $r_{\rm d}$ is fixed and $M_{\rm B}$ is set to two priors (SH0ES or M24), %, differing by 0.1 mag)
we observe that the DDR satisfies for the C19$^*$ subsample within $1-2\sigma$;
%M24 + C19$^*$ + FULL/GOLD satisfies the DDR within $1\sigma$, SH0ES + C19* +FULL/GOLD within $2\sigma$;
while the DDR deviate for the A20$^*$ subsample beyond $2\sigma$. %M24 + A20$^*$ +FULL within $2\sigma$. M24 + A20$^*$ +GOLD within $2.1\sigma$, and SH0ES + A20$^*$ +FULL/GOLD shows a significant deviation.
This indicates that a difference of $\sim 0.1 mag$ in $M_{\rm B}$ can lead to a deviation of about $\sim 1\sigma$ in the DDR.

By comparing the results obtained with the FULL/GOLD samples to those using only SNe ($z<3.6$) under the $P_1$ and $P_2$ models (Tab.~\ref{tab:tot}),
we find that the inclusion of GRB data generally shifts the best-fit values slightly toward smaller values with larger uncertainties on $\eta_0$. This is mainly because the current measurements of GRBs have intrinsically larger uncertainties than those of SNe Ia. % (see Appendix C for a brief discussion of these potential issues and their impact).
Nevertheless, GRBs remain valuable for extending DDR tests to higher redshifts, complementing the DDR constraints dominated by SNe Ia at low-redshift.
Moveover, some potential issues and limitations of GRB data, such as the luminosity standardization problem, might be impact on the DDR test results.
Furthermore, the choosing of the redshift division might affect the results of reconstruction of the data.
In Appendix B, we discuss the luminosity standardization problem with their impact on the test results.
%However, we have compared the results of reconstruction with redshift cutoffs at $z = 2.26$ and $z = 1.4$, and found
We find that that DDR tests with the Fermi data obtained by the simultaneous fitting method are consistent with ones of calibration at low-redshift at $z=1.4$.
We also briefly compare the DDR constraints derived with different choices of the redshift cutoff in Appendix B. We find the DDR constraints are insensitive to the choices of $z_{\rm cut}=1.4$ and $z_{\rm cut}=2.26$.

Due to the extensive use of various datasets, we have used two commonly parameterizations ($\eta(z) = 1 + \eta_0 z$) and $\eta(z) = 1 + \eta_0 z/(1+z)$). For different parameterizations can affect the results of DDR, we have compared other types of parameterized models, i.e., $\eta(z) = 1 + \eta_0 \ln(1+z)$ \cite{Nair2011}, $\eta(z) = (1+z)^{\eta_0}$ \cite{Holanda2017},  and found that the resulting DDR constraints are significantly unaffected. We have also compared our results for two commonly parameterizations with ones from recent result using only BAOs and SNe\footnote{It should be noted the inferred DDR constraints may partially inherit the calibration tension associated with $M_{\rm B}$ and $r_{\rm d}$, in which a $> 5\sigma$ discrepancy between the cosmological distance ladder built from BAO calibrated by the Planck/$\Lambda$CDM sound horizon and SNe Ia calibrated instead with the S$H_0$ES absolute magnitude assuming the DDR holds.
Conventional calibration of BAOs relies on estimation of $r_d$ from early universe observations by assuming a cosmological model.
Poulin et al.\cite{Poulin2024} emphasized the consequences of the cosmic calibration tension beyond the value of the Hubble constant $H_0$, and the implications for physics beyond $\Lambda$CDM.
%The `cosmic calibration tension' is a $> 5\sigma$ discrepancy between the cosmological distance ladder built from BAO calibrated by the Planck/$\Lambda$CDM sound horizon ($r_s$) and Type Ia supernovae (SN1a) calibrated instead with the S$H_0$ES absolute magnitude, assuming the DDR holds.
Shah et al.\cite{Shah2024a} used a novel deep learning framework (LADDER - Learning Algorithm for Deep Distance Estimation and Reconstruction)  with datasets Pantheon and BAOs for consistency checks, and calibration of high-redshift datasets such as GRBs.
Shah et al.\cite{Shah2024b} presented a recalibration of two independent BAO datasets (SDSS and DESI) by employing LADDER for model-independent estimation of $r_d$ and explore the impacts on $\Lambda$CDM cosmological parameters.
Kanodia et al.\cite{Kanodia2026} suggests that the observed departures from the DDR are more plausibly associated with early--late distance calibration differences related to the Hubble tension, rather than indicating a statistically significant violation of the DDR itself.}. %for the variability (shifts, accuracy, etc.) of the changes in DDR limitation results was thoroughly elaborated.
%Moreover, %in these two cases,
We find $\eta_0$ and $M_{\rm B}$ exhibit a negative correlation, which is consistent with \citet{Li2025}.
We also find that %in the cases of free  $M_{\rm B}$, %with $r_{\rm d} = 147.78$ and free $r_{\rm d}$,
the best-fit values of $M_{\rm B}$ are greater than or equal to $-19.45$ for all cases, which differs from the results of \citet{Li2025} using Pantheon+ combined with DESI DR2 by assuming a fixed $r_{\rm d} = 147.09\,\mathrm{Mpc}$, where the best-fit $M_{\rm B}$ is below $-19.5$ when $M_{\rm B}$ is free and $r_{\rm d} = 147.09~\mathrm{Mpc}$ is fixed, and the $1\sigma$ constraints on $M_{\rm B}$ were relatively broad (ranging from 0.2 to 0.79). In contrast, in our analysis the constraints on $M_{\rm B}$ are tighter, with the maximum $2\sigma$ range not exceeding
0.3.
%It should be noted that \citet{Kanodia2026} found $\eta(z) \simeq 1$ for a $\Lambda$CDM-consistent absolute magnitude $M_{\rm B} = -19.40$, while a deviation from unity appears when adopting the SH0ES calibration with $M_{\rm B} = -19.25$. This It
%Specifically, the (SN $\times M_{\rm B}^{\rm M24}$ + FULL + C19$^*$ + BAO) combination yields $\eta(z)$ consistent with unity within $1\sigma$, while the (SN $\times M_{\rm B}^{\rm SH0ES}$ + FULL + C19$^*$ + BAO) combination remains consistent with $\eta(z)=1$ at the $2\sigma$ level, showing a mild tendency toward deviation.
Our results based on the P1 and P2 models with fixing $r_{\rm d} = 147.09\,\mathrm{Mpc}$ are nearly consistent with those of Kanodia et al.\cite{Kanodia2026} by using Pantheon+ combined with DESI DR2 (%five data points,
excluding the $z \simeq 2.33$ point) with a fixed $r_{\rm d} = 147\,\mathrm{Mpc}$.
We emphasize that in the analysis where both the SN absolute magnitude $M_{\rm B}$ and the sound horizon scale $r_{\rm d}$ are fixed, the DDR test is no longer strictly model-independent.
%Fixing these two parameters effectively imposes strong external distance calibrations, which are known to exhibit non-negligible discrepancies among different determinations.
%As a result, the inferred DDR constraints may partially inherit the calibration tension associated with $M_{\rm B}$ and $r_{\rm d}$ \cite{Poulin2024,Shah2024a,Shah2024b}. %, which has been discussed for testing DDR \cite{Kanodia2026}.

\section{CONCLUSION AND DISCUSSIONS}\label{sec:conclu}

In this work, we independently reconstructed the luminosity distances of the Pantheon+ SNe Ia sample  ($0.01 < z \le 1.4$)  and the high-redshift GRBs ($1.4 < z < 8.2$) using ANNs
%\textbf{ANN}
approach with the ADDs $D_{\rm A}(z)$ %was non-parametrically reconstructed
from the DESI DR2 BAO data and SGLs to perform a model-independent test of the DDR.
For the absolute magnitude calibration, two priors were adopted:
$M_{\rm B}^{\rm M24} = -19.353^{+0.073}_{-0.078}~\mathrm{mag}$ and $M_{\rm B}^{\rm SH0ES} =  -19.253 \pm 0.027~\mathrm{mag}$.
%We tested the CDDR using both parametric and non-parametric approaches.
%In the parametric form, two parameterizations (P1 and P2) were considered, and three schemes were examined:
%(i) both $r_{\rm d}$ and $M_{\rm B}$ are free parameters;
%(ii) fixing $r_{\rm d} = 147.78~\mathrm{Mpc}$ (Planck value);
%(iii) fixing both $r_{\rm d} = 147.78~\mathrm{Mpc}$ and $M_{\rm B}^{\rm M24}$ or $M_{\rm B}^{\rm SH0ES}$.
In the parametric form, two parameterizations (P1 and P2) were considered, and two schemes were examined:
(i) fixing $r_{\rm d} = 147.09~\mathrm{Mpc}$ (Planck value);
(ii) fixing both $r_{\rm d} = 147.09~\mathrm{Mpc}$ and $M_{\rm B}^{\rm M24}$ or $M_{\rm B}^{\rm SH0ES}$.
The results show that, for all parameterizations in
%both
parametric reconstructions, We find that the fitting results using combinations with the FULL sample are consistent with the GOLD sample;
the C19 subsample slightly favoured the DDR over the A20 subsample.
%\textbf{And the GRB samples also has some limitations that should be noted. First, the measurement uncertainties of GRB spectra and energetics are relatively large, which increases the errors in the inferred distance moduli. Second, the inferred GRB distance moduli depend on the adopted empirical relations and calibration methods. These methodological choices can lead to variations in the DDR constraints.}
%the DDR shows significant deviations when combinations with $D_{\rm obs} > 1$ are not excluded.
%Moreover, %in these two cases,
%Specifically, for the C19$^*$ subsample, in case (i) and (ii), the DDR is both consistent within $1\sigma$.
%the deviation parameter $\eta_0$ remains consistent with $\eta_0 = 0$ within the $1\sigma$ confidence level.
%Moreover, we find a clear negative correlation between $M_{\rm B}$ and $\eta_0$;% is observed. Therefore,
%in case (ii), for the C19$^*$, we find that
%as well as a difference of 0.1 mag in $M_{\rm B}$ can lead to a deviation of about $1\sigma$ in the DDR.
%Moreover, a clear negative correlation between $M_{\rm B}$ and $\eta_0$ is observed, implying that a dimmer absolute magnitude tends to correspond to a larger best-fit $\eta_0$.
%These findings indicate that no statistically significant deviation from the DDR is detected within the precision of current data, supporting the validity of metric gravity, photon number conservation, and standard light propagation in the late-time Universe.
It should be noted that the inferred DDR constraints may partially inherit the calibration tension associated with $M_{\rm B}$ and $r_{\rm d}$. When both the SN absolute magnitude and the sound horizon scale are fixed, the analysis of DDR effectively imposes strong external distance calibrations.

Future work will focus on improving non-parametric reconstruction techniques using neural networks and extending this framework to constrain cosmological parameters %, the dark energy equation of state,
and the expansion history with next-generation probes such as LSST, \textit{Euclid}, SKA, and the Einstein Telescope. These upcoming surveys will enable DDR tests at sub-percent precision.%, offering powerful new avenues to explore fundamental physics and the nature of cosmic expansion.

\section*{Appendix A: ANN Architecture Selection and Training Validation}
To determine the optimal ANN architecture,
in this appendix, we performed a grid search over different network configurations, varying the number of hidden layers (\(N = 1 \text{--} 4\)) and the number of neurons per layer (\(2^n\), with \(5 \le n \le 8\)). All candidate architectures were trained for 6000 epochs using the Adam optimizer with an initial learning rate of \(10^{-3}\) and a weight decay of \(10^{-5}\). For Pantheon+, the total loss function consists of a \(\chi^2\) term and a KL divergence term, while for GRB data only the KL divergence is included. The architecture that minimizes the total loss was selected as the optimal ANN model.
The dataset was split into training and validation sets with an 80\%/20\% ratio to assess the generalization performance of the network.
During the ANN reconstruction, the reconstruction errors are incorporated into the loss function through the KL divergence term, which is optimized together with the reconstruction mean. The difference between the observed Gaussian distribution $P(\mu_\mathrm{obs}, \sigma_\mathrm{obs})$ and the ANN-predicted Gaussian distribution $Q(\mu_\mathrm{pred}, \sigma_\mathrm{pred})$ is measured by the KL divergence:
\begin{equation}
		D_{\mathrm{KL}}\left(P \parallel Q\right) = \frac{1}{2} \left[ \frac{\sigma_{\mathrm{obs}}^2}{\sigma_{\mathrm{pred}}^2} + \frac{(\mu_{\mathrm{pred}} - \mu_{\mathrm{obs}})^2}{\sigma_{\mathrm{pred}}^2} - 1 + \ln \frac{\sigma_{\mathrm{pred}}^2}{\sigma_{\mathrm{obs}}^2} \right].
\end{equation}

\begin{figure*}
	\centering
	% 单行两列
	\begin{minipage}{0.45\textwidth}
		\centering
		\includegraphics[width=\textwidth]{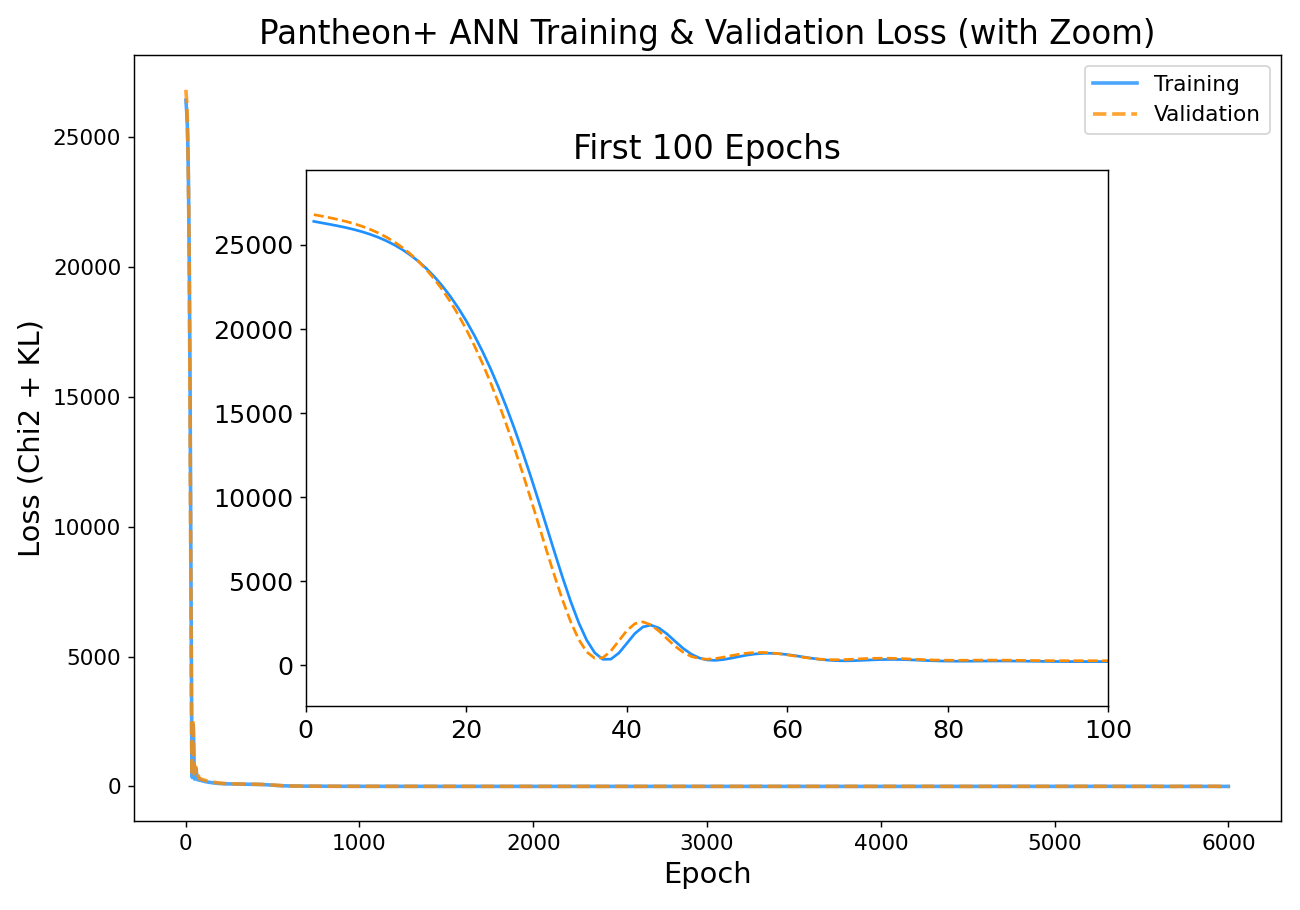}
		%		\subcaption{P1 model}
	\end{minipage}
	\hfill
	\begin{minipage}{0.45\textwidth}
		\centering
		\includegraphics[width=\textwidth]{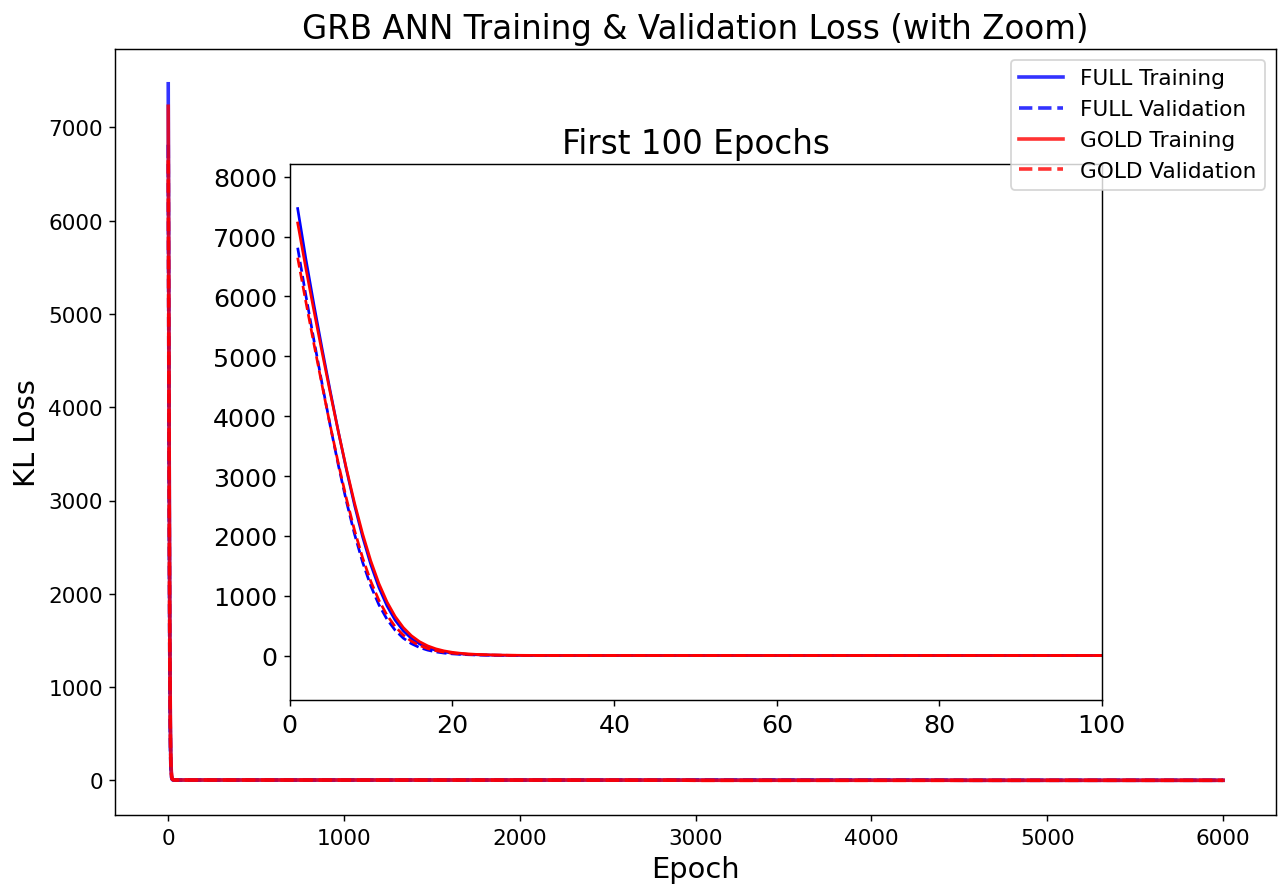}
		%		\subcaption{P2 model}
	\end{minipage}
	\captionsetup{singlelinecheck=off, justification=raggedright}
	\caption{
		Training/validation loss curves from the optimal network architecture for the Pantheon+ dataset (Left panel) and  %(main plot), details of the first 100 training epochs (inset plot).%, and loss function comparisons across different architectures (bottom-right inset). 		Right panel: Corresponding analysis for
the GRB dataset (Right panel).
	}
	\label{fig:loss}
\end{figure*}

Fig.~\ref{fig:loss} presents %the ANN architecture selection and
the training--validation results obtained using the \emph{optimal} ANN architecture simultaneously. The main panel shows the evolution of the training and validation losses as a function of epoch for the selected network, where the two curves nearly overlap throughout the entire training process, indicating no evident overfitting. The the central inset enlarges the first 100 epochs, demonstrating that the network rapidly converges and reaches a stable state at an early training stage.
%The lower-right inset compares the total losses obtained after 6000 epochs for different ANN architectures, from which the architecture with the minimum total loss is identified and adopted as the optimal model in this work.
These results confirm that the chosen ANN architecture exhibits good convergence and robustness, providing a reliable basis for the subsequent reconstruction analysis.

\begin{figure*}
	\centering
	% 一行四列
	\begin{minipage}[t]{0.45\textwidth} % 调整为四分之一宽度，留出一些间距
		\centering
		\includegraphics[width=\textwidth]{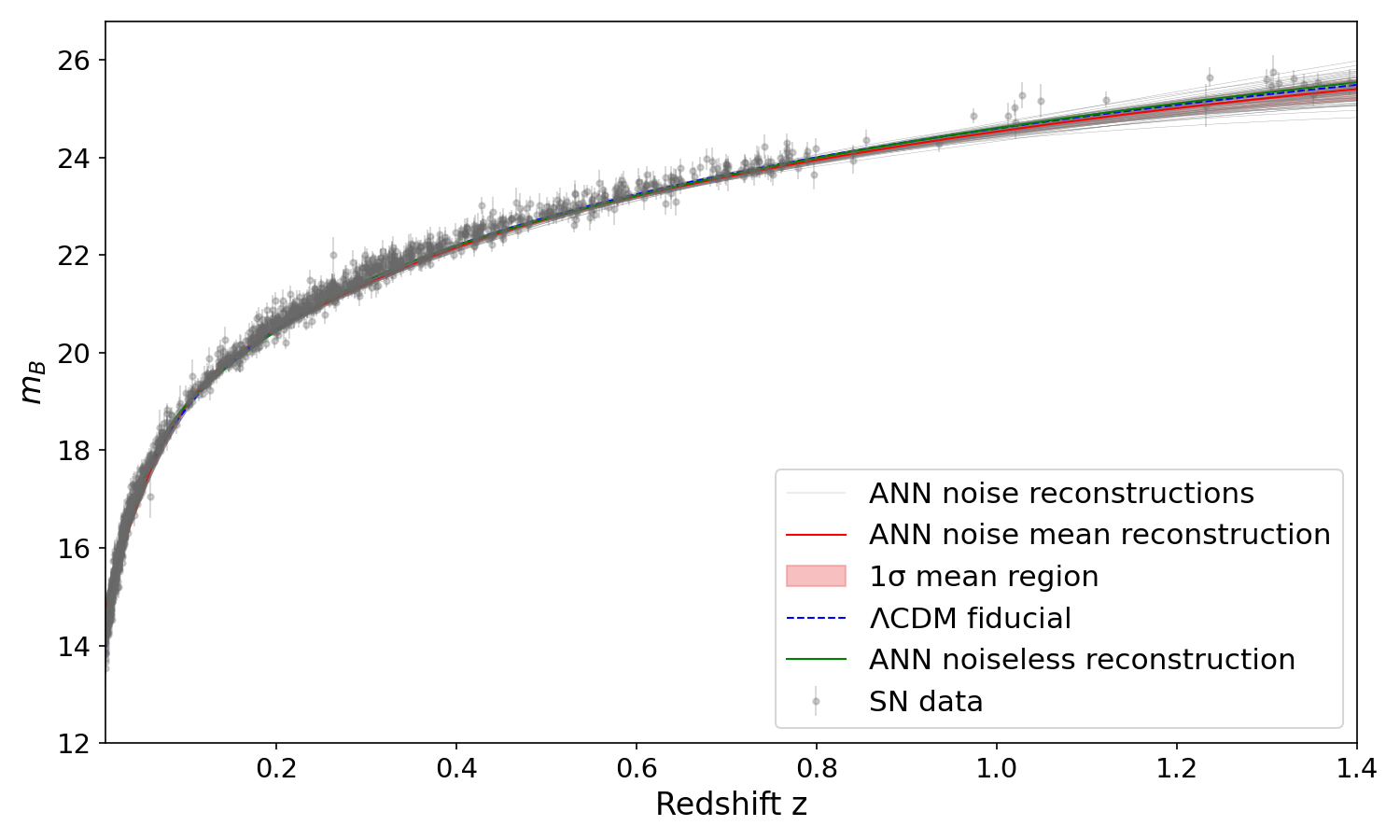}
		 \subcaption{Mock SN}
	\end{minipage}
	\hfill % 弹性水平间距
	\begin{minipage}[t]{0.45\textwidth}
		\centering
		\includegraphics[width=\textwidth]{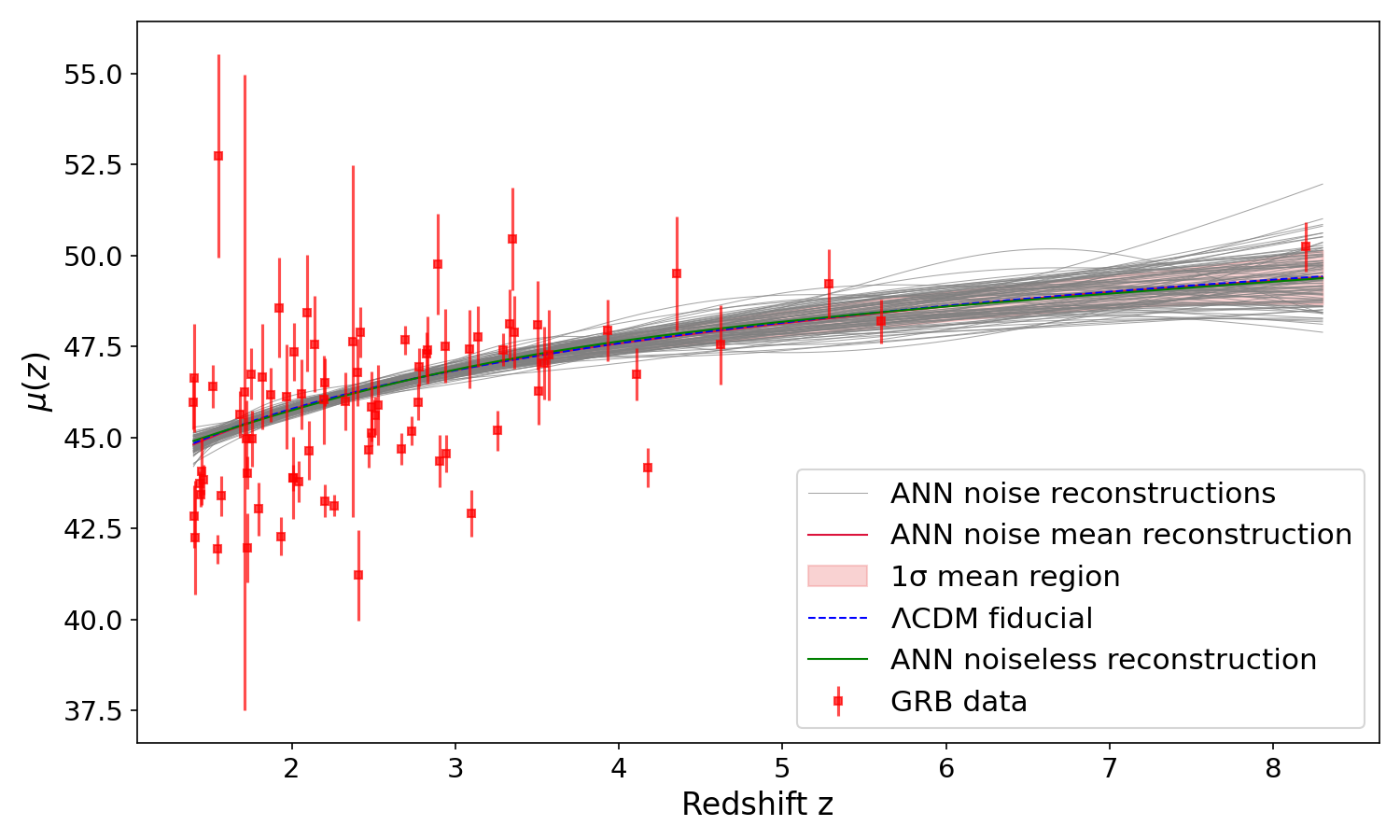}
		 \subcaption{Mock GRB}
	\end{minipage}
	
	\captionsetup{singlelinecheck=off, justification=raggedright}
%	\caption{Reconstruction  from mock data by the optimal ANN architecture. Panels (a) and (b) show the results of the mock SN and mock GRB samples, respectively. %where the networks are trained using our original optimal ANN architecture and loss function.
%}

     \caption{Reconstruction from mock data using the optimal ANN architecture.
     Panels (a) and (b) show the results for the mock SN and GRB samples, respectively.
     For the noiseless mock: the blue dashed line denotes the fiducial $\Lambda$CDM model, and the green line corresponds to the reconstruction from the noiseless mock sample. For the noisy mock:  the gray curves represent the ANN reconstructions obtained from 100 noisy realizations, the red line shows their mean reconstruction with the shaded region indicating the $1\sigma$ uncertainty.
     The GRB reconstruction from mock data is performed in the manner for the SN mock sample analogously.}
	\label{fig:mock}
\end{figure*}

With lower losses on their respective datasets, we find the optimal neural network has four fully connected layers with sizes [128, 128, 64, 32] for the Pantheon+ data, and three fully connected layers with sizes [128, 128, 128] for the \emph{Feimi} data.
%\section*{Appendix B: Validation with Mock Data}
In order to test the validation in ANN framework, we reconstruct by the optimal ANN from mock data  %to  test the DDR
%We validate the ANN reconstruction method and the DDR using mock data
generated according to the flat $\Lambda$CDM model fitting with Pantheon+ sample \cite{Brout2022} ($\Omega_M = 0.334$, $H_0 = 73.6\ \mathrm{km\ s^{-1}\ Mpc^{-1}}$) at the redshifts of the observational data.
%For both the SN and GRB mock samples, the uncertainty on the distance modulus is set as $\sigma^{\rm mock}_{\mu} = 0.02\mu_\mathrm{mock}$.
%For the SN sample, the apparent magnitude is first generated as
%$m_{\rm B}^\mathrm{mock} = \mu_\mathrm{mock} + M_{\rm B}$,
%with $M_{\rm B} = M_{\rm B}^{M24}$ fixed.
%The total uncertainty on the SN apparent magnitude is then
%$\sigma^{\rm mock}_{m_{\rm B}} = \sigma^{\rm mock}_{\mu}$.
%For the FULL GRB sample, the uncertainty on the mock distance modulus is $\sigma_\mu^\mathrm{mock} = 0.02\mu_\mathrm{mock}$.
For the SN mock sample, the uncertainties of mock data can be generated by two ways:
(i) Noiseless mock: the fiducial data values are taken directly from the $\Lambda$CDM prediction at the observed redshifts, i.e., $\mu_\mathrm{mock} = \mu_{\Lambda\mathrm{CDM}}(z_\mathrm{obs})$ and $m_{B,\mathrm{fid}}^\mathrm{mock} = \mu_\mathrm{mock} + M_B$ with $M_{\rm B}=M_{\rm B}^{M24}$ fixed.
The original observational covariance $\mathbf{C}_\mathrm{obs}$ for the mock sample are taken account into $\chi^2$ during training.
(ii) Noisy mock: we draw the uncertainties from $\sim \mathcal{N}(0, \mathbf{C}_\mathrm{obs})$ to generate the noise of mock sample: $m_B^\mathrm{mock} = m_{B,\mathrm{fid}}^\mathrm{mock} + \mathrm{noise}$. We generate 100 realizations and perform the ANN reconstruction for each sample. The original observational covariance $\mathbf{C}_\mathrm{obs}$ is also used during training.
For the GRB mock sample, the mock data are generated analogously.
The results of reconstruction are shown in Fig.~\ref{fig:mock}. %where panels (a) and (b) display the ANN reconstruction results for the mock SN and mock GRB samples, respectively. %and panels (c) and (d) show the DDR tests for the P1 and P2 models based on the mock data.}

We have tested different ANN architectures, multiple random seeds and activation functions settings to find that the reconstructed $m_{\rm B}(z)$ or $\mu(z)$ fluctuate significantly less than the observational uncertainties. In Tab~\ref{tab:hyp},
we show the DDR constraints under different neural network configurations for the dataset combination SN+FULL+C19*+BAO, which indicates that the results of DDR are NOT sensitive to initialization and architectural choices.

\begin{table*}
	\centering
	\tiny
	\captionsetup{singlelinecheck=off, justification=raggedright}
			\caption{\label{tab:hyp}
		Fitting results of the $P_1$ and $P_2$ models($1\sigma$ and $2\sigma$ confidence levels) under different neural network configurations for the dataset combination SN+FULL+C19*+BAO.
		The optimal neural network architecture: [128, 128, 64, 32] for Pantheon+ and [128, 128, 128] for \emph{Feimi} FULL sample with random seed used for network initialization: $r_0=68$, and activation function: SiLU.
		For comparison, we choose neural network architectures [128, 128, 128] for SNe and [256,256] for GRBs with relatively low losses on their respective datasets, random seed: $r_1=60,r_2=80$ and activation function: ELU \cite{Clevert2015}.
		%The symbols $r_0$, $r_1$, and $r_2$ denote different random seeds used for network initialization (with $r_0=68$, $r_1=60$, and $r_2=80$), and SiLU or ELU \cite{Clevert2015} are activation functions.
		%Optimal ANN in This work=SN:[128,128,64,32] \& GRB:[128,128,128]+$r_0$+SiLU;
		%For comparison, we choose neural network architectures [128, 128, 128] for SNe and [256, 256] for GRBs with relatively low losses on their respective datasets.
		For ANN reconstructions with SNe and FULL samples, we use full-batch training with 6000 epochs.
	}
	\renewcommand{\arraystretch}{1.8}
	\setlength{\tabcolsep}{4pt}
	\resizebox{\textwidth}{!}{%
		\begin{tabular}{l cccc}
			\hline
			\hline
			\multirow{2}{*}{Configuration}
			& \multicolumn{2}{c}{$P_1$ model} & \multicolumn{2}{c}{$P_2$ model} \\
			\cmidrule(lr){2-3} \cmidrule(lr){4-5}
			& $\eta_0$ & $M_{\rm B}$ & $\eta_0$ & $M_{\rm B}$ \\
			\hline
			Optimal(This work)
			& $0.01^{+0.05+0.11}_{-0.05-0.09}$ & $-19.45^{+0.11+0.22}_{-0.12-0.23}$
			& $0.01^{+0.13+0.26}_{-0.12-0.23}$ & $-19.44^{+0.13+0.27}_{-0.14-0.28}$ \\
			
			A1 (SN: [128,128,128])
	     	& $0.02^{+0.05+0.10}_{-0.05-0.09}$ &
	        $-19.45^{+0.11+0.21}_{-0.11-0.23}$
		    & $0.02^{+0.13+0.26}_{-0.12-0.22}$ & $-19.44^{+0.14+0.26}_{-0.14-0.28}$ \\
			
			A2 (GRB: [256,256])
			& $0.00^{+0.05+0.10}_{-0.04-0.08}$ & $-19.55^{+0.11+0.21}_{-0.11-0.23}$
			& $-0.03^{+0.12+0.25}_{-0.11-0.22}$ & $-19.53^{+0.13+0.26}_{-0.14-0.28}$ \\
			
			A3 ($r_1=60$)
			& $0.01^{+0.05+0.10}_{-0.05-0.09}$ & $-19.46^{+0.11+0.21}_{-0.11-0.23}$
			& $0.00^{+0.13+0.26}_{-0.12-0.22}$ & $-19.43^{+0.13+0.26}_{-0.14-0.28}$ \\
			
			A4 ($r_2=80$)
			& $0.02^{+0.05+0.10}_{-0.05-0.09}$ & $-19.45^{+0.11+0.21}_{-0.11-0.23}$
			& $0.01^{+0.13+0.26}_{-0.12-0.22}$ & $-19.44^{+0.13+0.26}_{-0.14-0.28}$ \\
			
			A5 (ELU)
			& $0.02^{+0.05+0.10}_{-0.05-0.09}$ & $-19.44^{+0.11+0.21}_{-0.11-0.23}$
			& $0.02^{+0.12+0.26}_{-0.11-0.22}$ & $-19.43^{+0.13+0.26}_{-0.13-0.27}$ \\
			
			\hline
			\hline
		\end{tabular}
	}
%	\vspace{2mm}
	\footnotesize
	\noindent
	\raggedright
	\textit{Notes.}
%	Optimal (This work)=SN:[128,128,64,32] \& GRB:[128,128,128]+$r_0$+SiLU;
	A1=SN:[128,128,128] \& GRB:[128,128,128]+$r_0$+SiLU;
	A2=SN:[128,128,64,32] \& GRB:[256,256]+$r_0$+SiLU;
	A3=SN:[128,128,64,32] \& GRB:[128,128,128]+$r_1$+SiLU;
	A4=SN:[128,128,64,32] \& GRB:[128,128,128]+$r_2$+SiLU;
	A5=SN:[128,128,64,32] \& GRB:[128,128,128]+$r_0$+ELU.
	
\end{table*}

\section*{Appendix B: Impact of GRBs calibration and Redshift Cutoff on the DDR Tests}

Finally, we discuss the impact on the DDR of GRBs calibration and the redshift cutoff. %, which used to separate the low-redshift and high-redshift samples
%in Tab~\ref{tab:zcut}.
We calibrate the \emph{Fermi} data by the simultaneous fitting method\footnote{For the luminosity standardization problem, recent findings suggest that the GRB relation parameters remain consistent by the simultaneous fitting method in which the parameters of cosmological models and the GRB relation parameter  are fitted simultaneously, implying that GRB data can be standardized across different cosmological models within error margins \citep{Khadka_2021}.} under the flat $\Lambda$CDM model. %DDR tests with the \emph{Fermi} data at $z=1.4$ obtained by the simultaneous fitting method are summarized in Tab~\ref{tab:zcut}.
We find that DDR tests with the \emph{Fermi} data at $z>1.4$ obtained by the simultaneous fitting method are consistent with ones %from the \emph{Fermi} data at $z>1.4$
by calibrating from low-redshift sample at $z<1.4$. Fitting results of DDR with GRBs with SN + C19* + BAO by different calibration methods are summarized in Tab~\ref{tab:zcut}.
%The uncertainties of GRB data are relatively large, mainly due to the luminosity correlations and observational errors, which increase the uncertainty of $\eta_0$ in DDR tests (see Section ~\ref{sec:test_ddr}). Nevertheless, GRBs remain valuable for extending DDR tests to higher redshifts.
To further investigate the impact of $z_{\rm cut}$ %, which used to separate the low-redshift and high-redshift samples,
on the calibration parameters and the DDR parameter $\eta_0$, we adopt $z_{\rm cut}=2.26$ as the redshift division and directly compare DDR results to ones obtained by  $z_{\rm cut}=1.4$ with %these two choices using
the same data combinations.
It should be note that the uncertainties of the GRB calibration parameters propagate into the luminosity distance $D_{\rm L}$, and subsequently enter the uncertainty of $\eta_0$.
We find that the minor variations in the calibration parameters between low- and high-redshift sub-samples by $z_{\rm cut}=2.26$ have only a limited effect on the reconstructed luminosity distances. %and the resulting DDR constraints.}
Fitting results of DDR with GRBs with SN + C19* + BAO by different redshift cutoff are summarized in Tab~\ref{tab:zcut}).
We find results with $z_{\rm cut}=2.26$ and $z_{\rm cut}=1.4$ are consistent each other; %from the \emph{Fermi} data at $z>1.4$
and a slight positive shift of the central values of $\eta_0$ with $z_{\rm cut}=2.26$ compared to the ones obtained with $z_{\rm cut}=1.4$.
\begin{table*}
	\centering
	\tiny
	\captionsetup{singlelinecheck=off, justification=raggedright}
	\caption{\label{tab:zcut}Fitting results of the $P_1$ and $P_2$ models($1\sigma$ and $2\sigma$ confidence levels) by free $M_{\rm B}$, using the simultaneous fitting method and redshift cutoff $z_{\rm cut}=2.26$. All results are obtained using the combined dataset SN+FULL/GOLD+C19*+BAO. For the simultaneous fitting method, we use the FULL(151 data) and GOLD(123 data) GRB samples jointly fitted with the Amati relation parameters ($a$, $b$, $\sigma_{\rm int}$), as well as $H_0$ and $\Omega_{\rm m}$ under the $\Lambda$CDM cosmological model assumption. SNe at $z<1.4$ are use to the reconstructed $m(z)$ and GRBs at $z\ge1.4$ are use to the reconstructed $\mu(z)$, respectively. %Following \citet{WangLiang2024}, in the low-redshift calibration method, GRBs are calibrated using the Amati relation based on low-redshift ($z<1.4$) supernovae, and the calibrated relation is then applied to reconstruct the $\mu(z)$ of high-redshift GRBs.
For the low-redshift calibration method with redshift cutoff $z_{\rm cut}=2.26$, we use SNe at $z<2.26$ %to the reconstructed $m(z)$
and GRBs at $z>2.26$ and for the reconstructions, respectively. % and different supernova absolute magnitude calibrations (M24 and SH0ES).
%In the second and third rows, the GRB reconstruction employs the same ANN architecture as that used in this work for the FULL/GOLD samples.
}
	\renewcommand{\arraystretch}{1.8}
	\setlength{\tabcolsep}{4pt}
    \resizebox{\textwidth}{!}{%
%	\begin{tabular}{l l cc cc}
%		\hline
%		\hline
%		\multirow{2}{*}{Method} & \multirow{2}{*}{GRB Sample} & \multicolumn{2}{c}{$P1$ model} & \multicolumn{2}{c}{$P2$ model} \\
%		\cmidrule(lr){3-4} \cmidrule(lr){5-6}
%		& & $\eta_0$ & $M_B$ & $\eta_0$ & $M_B$ \\
%		\hline
%		
%		\multirow{2}{*}{The simultaneous fitting method}
%		& FULL & $0.01^{+0.04+0.09}_{-0.04-0.07}$ & $-19.59^{+0.10+0.19}_{-0.11-0.21}$
%		& $0.00^{+0.11+0.23}_{-0.11-0.20}$ & $-19.57^{+0.12+0.23}_{-0.12-0.24}$ \\
%		
%		& GOLD & $0.01^{+0.04+0.08}_{-0.04-0.07}$ & $-19.58^{+0.10+0.19}_{-0.10-0.21}$
%		& $0.00^{+0.11+0.23}_{-0.10-0.19}$ & $-19.58^{+0.12+0.23}_{-0.12-0.25}$ \\
%		
%		\hline
%		
%		\multirow{2}{*}{The low-redshift calibration method($z_{\rm cut}=2.26$)}
%		& FULL & $0.08^{+0.04+0.09}_{-0.04-0.08}$ & $-19.55^{+0.11+0.21}_{-0.11-0.22}$
%		& $0.13^{+0.13+0.26}_{-0.12-0.22}$ & $-19.54^{+0.14+0.26}_{-0.13-0.27}$ \\
%		
%		& GOLD & $0.06^{+0.04+0.08}_{-0.04-0.07}$ & $-19.52^{+0.10+0.19}_{-0.10-0.21}$
%		& $0.11^{+0.12+0.24}_{-0.11-0.20}$ & $-19.52^{+0.12+0.24}_{-0.13-0.25}$ \\
%		
%		\hline
%		
%		\multirow{2}{*}{The low-redshift calibration method($z_{\rm cut}=1.4$, Tab~\ref{tab:tot})}
%		& FULL & $0.01^{+0.05+0.11}_{-0.05-0.09}$ & $-19.45^{+0.11+0.22}_{-0.12-0.23}$
%		& $0.01^{+0.13+0.26}_{-0.12-0.23}$ & $-19.44^{+0.13+0.27}_{-0.14-0.28}$ \\
%		
%		& GOLD & $-0.02^{+0.05+0.10}_{-0.04-0.08}$ & $-19.40^{+0.11+0.21}_{-0.11-0.23}$
%		& $-0.05^{+0.12+0.25}_{-0.11-0.21}$ & $-19.38^{+0.13+0.26}_{-0.13-0.27}$ \\
%		
%		\hline
%		\hline
%	\end{tabular}
    \begin{tabular}{l l cc cc}
	\hline
	\hline
	\multirow{2}{*}{Method} & \multirow{2}{*}{GRB Sample} & \multicolumn{2}{c}{$P_1$ model} & \multicolumn{2}{c}{$P_2$ model} \\
	\cmidrule(lr){3-4} \cmidrule(lr){5-6}
	& & $\eta_0$ &  & $\eta_0$ &  \\
	\hline
	
	\multirow{2}{*}{The simultaneous fitting method}
	& FULL & $0.01^{+0.04+0.09}_{-0.04-0.07}$ &  & $0.00^{+0.11+0.23}_{-0.10-0.20}$ &  \\
	
	& GOLD & $0.01^{+0.04+0.08}_{-0.04-0.08}$ &  & $0.01^{+0.11+0.23}_{-0.10-0.19}$ &  \\
	
	\hline
	
	\multirow{2}{*}{The low-redshift calibration method($z_{\rm cut}=2.26$)}
	& FULL & $0.08^{+0.04+0.09}_{-0.04-0.08}$ &  & $0.13^{+0.13+0.26}_{-0.12-0.22}$ &  \\
	
	& GOLD & $0.06^{+0.04+0.08}_{-0.04-0.07}$ &  & $0.11^{+0.11+0.23}_{-0.11-0.21}$ &  \\
	
	\hline
	
	\multirow{2}{*}{The low-redshift calibration method($z_{\rm cut}=1.4$, Tab~\ref{tab:tot})}
	& FULL & $0.01^{+0.05+0.11}_{-0.05-0.09}$ &  & $0.01^{+0.13+0.26}_{-0.12-0.23}$ &  \\
	
	& GOLD & $-0.02^{+0.05+0.10}_{-0.04-0.08}$ &  & $-0.05^{+0.12+0.25}_{-0.11-0.21}$ &  \\
	
	\hline
	\hline
\end{tabular}
}
\end{table*}

\section*{Appendix C: Robustness Tests on the treating of the sound horizon scale $r_{\mathrm{d}}$}
\begin{figure*}
	\centering	
	\begin{minipage}{0.4\textwidth}
		\centering
		\includegraphics[width=\textwidth]{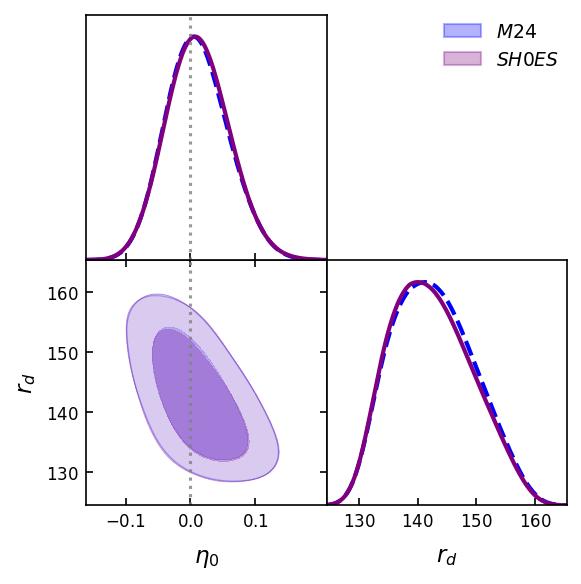}
		\subcaption{P1 model}
	\end{minipage}
	\hfill
	\begin{minipage}{0.4\textwidth}
		\centering
		\includegraphics[width=\textwidth]{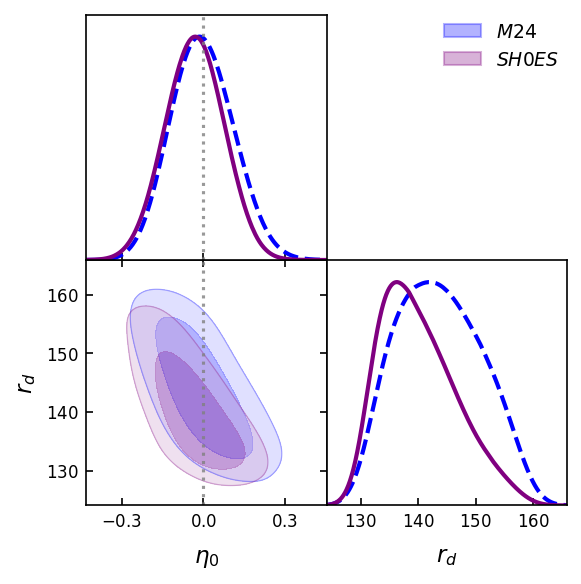}
		\subcaption{P2 model}
	\end{minipage}

	\captionsetup{singlelinecheck=off, justification=raggedright}
    \caption{Constraints on the cosmological parameters ($\eta_0$, $r_{\rm d}$) using SN$\times M_{\rm B}^{\rm SH0ES}$/SN$\times M_{\rm B}^{\rm M24}$+FULL+C19$^*$+BAO, with $r_{\rm d}$ treated as a free parameter. Panels (a) and (b) correspond to the P1 and P2 models, respectively.}
	%	\caption{Constraints on the cosmological parameters $\eta_0$, $M_{\rm B}$ using SGLs+BAOs with SNe Ia+GRBs. The left panel shows results for the P1 model, and the right panel shows results for the P2 model.}
	\label{fig:six}
\end{figure*}

To assess the impact of fixing the sound horizon scale $r_d$ in the DDR analysis, we perform additional robustness tests at the likelihood level.
We treat $r_{\rm d}$ as a free parameter with a prior range $r_{\rm d} \in [130,160]$ Mpc, and fixing $M_{\rm B}$ (M24 or SH0ES) to each calibration. The joint posterior distributions of $(\eta_0, r_d)$ with SN+FULL+C19$^*$+BAO are shown in Fig.~\ref{fig:six},
%With the dataset FULL+BAO+$C19^*$ for P1 model: (i) $M_{\rm B}=M_{\rm B}^{\mathrm{M24}}$: $\eta_0 = 0.01^{+0.05}_{-0.05}$, $r_{\rm d} = 142.23^{+7.94}_{-7.22}\ \mathrm{Mpc}$;
%(ii) $M_{\rm B}=M_{\rm B}^{\mathrm{SH0ES}}$: $\eta_0 = 0.01^{+0.05}_{-0.05}$, $r_{\rm d} = 141.77^{+8.09}_{-6.90}\ \mathrm{Mpc}$;
%for P2 model: (i) $M_{\rm B}=M_{\rm B}^{\mathrm{M24}}$: $\eta_0 = -0.01^{+0.12}_{-0.11}$, $r_{\rm d} = 143.29^{+8.75}_{-8.00}\ \mathrm{Mpc}$;
%(ii) $M_{\rm B}=M_{\rm B}^{\mathrm{SH0ES}}$: $\eta_0 = -0.03^{+0.11}_{-0.11}$, $r_{\rm d} = 138.92^{+8.13}_{-6.04}\ \mathrm{Mpc}$.
which are  summarized in Tab~\ref{tab:rd_err} (CASE A). 
We find a strong negative correlation between $r_{\rm d}$ and $M_B$.
In addition, for the branches where $r_d$ is fixed, we propagate the uncertainty ($r_d = 147.09 \pm 0.26$ Mpc) into the testing of $\eta_0$. The results with SN+FULL+C19$^*$+BAO and fixing $M_{\rm B}$ (M24) are summarized in Tab~\ref{tab:rd_err} (CASE B).
We find the impact of the uncertainty of $r_d$ on the DDR constraints is  negligible. %from corresponding comparison cases.

\begin{table*}
	\centering
	\tiny
	\captionsetup{singlelinecheck=off, justification=raggedright}
	\caption{\label{tab:rd_err}
		Fitting results of the $P_1$ and $P_2$ models ($1\sigma$ and $2\sigma$ confidence levels) with SN+FULL+C19$^*$+BAO by treating $r_d$ free and propagating the $r_d$ uncertainty on the DDR constraints.}
	
	\renewcommand{\arraystretch}{1.8}
	\setlength{\tabcolsep}{4pt}
	
	\resizebox{\textwidth}{!}{%
		\begin{tabular}{l ccc ccc}
			\hline
			\hline
			\multirow{2}{*}{Dataset}
			& \multicolumn{3}{c}{$P_1$ model} & \multicolumn{3}{c}{$P_2$ model} \\
			\cmidrule(lr){2-4} \cmidrule(lr){5-7}
			& $\eta_0$ & $M_{\rm B}$ & $r_{\rm d}$
			& $\eta_0$ & $M_{\rm B}$ & $r_{\rm d}$ \\
			\hline
			
			CASE A
			& $0.01^{+0.05+0.10}_{-0.05-0.10}$  &  $M_{\rm B}^{\rm M24} = -19.353^{+0.073}_{-0.078}$ & $142.23^{+7.94}_{-7.22}$
			& $-0.01^{+0.12+0.24}_{-0.11-0.22}$ & $M_{\rm B}^{\rm M24} = -19.353^{+0.073}_{-0.078}$  & $143.29^{+8.75}_{-8.00}$ \\

			& $0.01^{+0.05+0.10}_{-0.05+0.10}$  & $M_{\rm B}^{\rm SH0ES} =  -19.253 \pm 0.027$ & $141.77^{+8.09}_{-6.90}$
			& $-0.03^{+0.11+0.22}_{-0.11-0.22}$ & $M_{\rm B}^{\rm SH0ES} =  -19.253 \pm 0.027$ & $138.92^{+8.13}_{-6.04}$ \\
			
			\hline
			%
%			C
%			& $0.01^{+0.05+0.11}_{-0.05-0.09}$ & $-19.45^{+0.11+0.22}_{-0.12-0.23}$ & --
%			& $0.01^{+0.13+0.26}_{-0.12-0.23}$ & $-19.44^{+0.13+0.27}_{-0.14-0.28}$ & -- \\
%			
%			D
%			& $0.01^{+0.05+0.11}_{-0.05-0.09}$ & $-19.45^{+0.11+0.22}_{-0.12-0.24}$ & --
%			& $0.01^{+0.13+0.26}_{-0.12-0.23}$ & $-19.43^{+0.13+0.27}_{-0.14-0.28}$ & -- \\
%			
			\hline
			
			CASE B
			& $-0.01^{+0.04+0.08}_{-0.04-0.07}$ & $M_{\rm B}^{\rm M24} = -19.353^{+0.073}_{-0.078}$ & $r_d = 147.09$
			& $-0.05^{+0.08+0.17}_{-0.08-0.16}$ & $M_{\rm B}^{\rm M24} = -19.353^{+0.073}_{-0.078}$ & $r_d = 147.09$ \\

			& $-0.01^{+0.04+0.08}_{-0.04-0.07}$ & $M_{\rm B}^{\rm M24} = -19.353^{+0.073}_{-0.078}$ & $r_d = 147.09 \pm 0.26$
			& $-0.04^{+0.08+0.17}_{-0.08-0.16}$ & $M_{\rm B}^{\rm M24} = -19.353^{+0.073}_{-0.078}$ & $r_d = 147.09 \pm 0.26$ \\
			
			\hline
			\hline
		\end{tabular}%
	}
	
	\footnotesize
	\raggedright
	
	\textit{Notes.}
	CASE A: Fitting results with SN+FULL+C19$^*$+BAO by treating $r_d$ free and fixing $M_{\rm B}$(M24 or SH0ES).
	%C: SN+FULL+C19$^*$+BAO (this work, without $\sigma_{r_d}$).
%	D: SN+FULL+C19$^*$+BAO$^e$ (including $r_d$ uncertainty).
	CASE B: Fitting results with SN+FULL+C19$^*$+BAO by propagating the $r_d$ uncertainty and and fixing $M_{\rm B}$(M24).
	The propagated uncertainty of $\eta_{\rm obs}$ is
	$\sigma_{\eta_{\rm obs}}^2 = \eta_{\rm obs}^2 \left[
	\left( \frac{\sigma_{D_{\rm L}}(z)}{D_{\rm L}(z)} \right)^2 +
	\left( \frac{\sigma_{D_{\rm A}}(z)}{D_{\rm A}(z)} \right)^2 +
	\left( \frac{\sigma_{r_d}}{r_d} \right)^2 \right]$.
	
\end{table*}

\section*{ACKNOWLEDGMENTS}
This project was supported by the Guizhou Provincial Science and Technology Foundation:
QKHJC-ZK[2021] Key 020 and QKHJC-ZK[2024] general 443. P. Wu was supported by the National Natural Science Foundation of China (Grant No. 12275080). X. Fu was supported by the National Natural Science Foundation of China under Grants No. 12375045.

\section*{DATA AVAILABILITY}
Data are available at the following references:
the \emph{Fermi} GRB sample from \cite{WangLiang2024},
the DESI DR2 BAO data from \cite{DESI2025a,DESI2025b},
SGLs data from \cite{Chen2019} and \cite{Amante2020},
as well as the Pantheon+ compilation from \cite{Scolnic2022}.

\bibliography{biblio3}

\end{document}